\documentclass{emulateapj}
\usepackage{epsfig}
\usepackage{apjfonts}
\usepackage{epsfig}

\newcommand{\lsim}{\lower0.6ex\vbox{\hbox{$ \buildrel{\textstyle
        <}\over{\sim}\ $}}}
\newcommand{\gsim}{\lower0.6ex\vbox{\hbox{$ \buildrel{\textstyle
        >}\over{\sim}\ $}}}
\newcommand{\hmpc}{ h^{-1}{\rm Mpc}}\newcommand{\Dvir}{ \Delta_{\rm vir}}
\newcommand{\rhomean}{\rho_{\rm ave}}
\newcommand{\Vmax}{V_{\rm max}}
\newcommand{\Vin}{V_{\rm in}}
\newcommand{\Vnow}{V_{\rm now}}

\newcommand{\Mvir}{M_{\rm vir}}
\newcommand{\Nc}{N_{c}}
\newcommand{\hkpc}{ h^{-1}{\rm kpc}}
\newcommand{\Msun}{M_{\odot}}
\newcommand{\hMsun}{h^{-1} \Msun}
\newcommand{\hMpc}{h^{-1} {\rm Mpc}}

\newcommand{\lcdm}{$\Lambda$CDM~}
\newcommand{\kms}{km s$^{-1}$}
\newcommand{\NN}{\left<N(N-1)\right>}
\newcommand{\Mpc}{\rm Mpc}
\newcommand{\N}{N(M)}
\newcommand{\Ns}{N_{s}(M)}
\newcommand{\Ncen}{N_{c}(M)}

\newcommand{\Rvir}{R_{\rm vir}}
\newcommand{\beq}{\begin{equation}}
\newcommand{\eeq}{\end{equation}}

\shorttitle{Close Galaxy Counts and Hierarchical Structure Formation}
\shortauthors{Berrier et~al.}

\begin{document}

\title{Close Galaxy Counts as a Probe of Hierarchical Structure Formation
}
\author{
Joel C. Berrier,
James S. Bullock,
Elizabeth J. Barton, 
Heather D. Guenther
}
\affil{Center for Cosmology, Department of Physics and Astronomy,
The University of California at Irvine, Irvine, CA 92697, USA}
\author{Andrew R. Zentner,
Risa H. Wechsler\altaffilmark{1},
}
\affil{Kavli Institute for Cosmological Physics and 
Department of Astronomy and Astrophysics, The University of Chicago, 
Chicago, IL 60637, USA}
\altaffiltext{1}{Hubble Fellow}

\begin{abstract}

Standard \lcdm predicts that the major merger rate of galaxy-size dark
matter  halos rises  rapidly  with increasing  redshift.  The  average
number of  close companions per galaxy,  ${\rm N_c}$ is  often used to
infer the  galaxy merger  rate.  Recent observational  studies suggest
that  ${\rm  N_c}$ evolves  very  little  with  redshift, in  apparent
conflict  with theoretical expectations  for the  merger rate  of dark
matter  halos.  We  use  a "hybrid"  N-body  simulation plus  analytic
substructure model  to build  theoretical galaxy catalogs  and measure
${\rm N_c}$  in the same  way that it  is measured in  observed galaxy
samples.  When we identify dark matter subhalos with galaxies, we show
that the observed lack of close pair count evolution arises naturally.
This unexpected result  is caused by the fact  that there are multiple
subhalos (galaxies) per host dark matter halo, and observed pairs tend
to reside in  massive halos that contain several  galaxies.  There are
fewer  massive host halos  at early  times and  this dearth  of galaxy
groups at high redshift compensates  for the fact that the merger rate
{\em per host halo} is higher  in the past.  We compare our results to
$z \sim  1$ DEEP2 data and  to $z \sim  0$ data that we  have compiled
from  the SSRS2  and the  UZC  redshift surveys.   The observed  close
companion counts match our  simulation predictions well, provided that
we  assume  a monotonic  mapping  between  galaxy  luminosity and  the
maximum circular velocity of each  subhalo in our catalogs at the time
when it was first accreted onto its host halo.  This strongly suggests
that satellite galaxies are  significantly more resilient to mass loss
than are dissipationless dark  matter subhalos. Finally, we argue that
while ${\rm N_c}$ does not provide a direct measure of the merger rate
of host  dark matter  halos, it offers  a powerful means  to constrain
both the Halo Occupation  Distribution and the spatial distribution of
galaxies  within dark matter  halos.  Interpreted  in this  way, close
pair counts  provide a  useful test of  galaxy formation  processes on
$\sim$10-100 kpc scales.

\end{abstract}



\section{Introduction} \label{sec:intro}

Close pairs of galaxies  have provided an important empirical platform
for evaluating  theories of galaxy  and structure formation  since the
early   studies   of   \citet{Holmberg1937}.    Within   the   modern,
hierarchical  picture  of  cosmological  structure  formation,  galaxy
mergers     are      expected     to     be      relatively     common
\citep[e.g.][]{Blumenthal84,LC93,kolatt:99,Gottlober01,maller:05}   and
the  enumeration of  close galaxy  pairs or  morphologically disturbed
systems  has  long   been  used  to  probe  the   galaxy  merger  rate
\citep[][]{z&k89, Burkey94,  Carlberg94, Woods95, Y&E95,  P97, Neus97,
Carlberg2000,   LeFevre2000, P2002, Conselice03, Bundy04,   masjedi05,
Belletal06, Lotz06}.  Comparisons  of this kind are clearly important,
as observed  correlations  between  galaxy  characteristics  and their
environments suggest that  interactions   play an essential   role  in
setting   many galaxy  properties       \citep[e.g.][]{ToomreToomre72,
LarsonTinsley78, Dressler80, PG84, Barton:00, Barton:03}.

While traditional merger rate estimates have provided empirical tests
of general aspects of the formation of galaxies, they have been less
useful in testing specific models of galaxy formation or in
constraining the background cosmological model.  One of the primary
difficulties lies in connecting theoretical predictions with
observational data.  For example, the standard approach uses the
observed close-pair count (or similarly, an observed
morphologically-disturbed galaxy count) to derive a merger rate
through approximate merger timescale arguments.  This {\em inferred}
merger rate is then compared to theoretical expectations for {\em
  field} dark matter halo merger rates, which themselves are quite
sensitive to the precise masses and mass ratios of halos under
consideration \citep[ see] [for a discussion of predicted {\em
    galaxy} merger rates] {maller:05,Belletal06}.  In this paper we
exploit the information contained in close-companion counts by
adopting a qualitatively different approach.  We use a ``hybrid''
analytic plus numerical $N$-body model to predict close-companion
counts directly and to compare these predictions with observed
companion counts.  Encouragingly, the results from this method reproduce
observed trends with redshift and number density in galaxy companion
counts.  More generally, our results suggest that while close
companion counts are only indirectly related to dark matter halo
merger rates, they may be used directly to constrain the number of
galaxy pairs per {\em host} dark matter halo.  In this way, companion
counts provide an important general constraint on the occupation of
halos by galaxies and, thereby, on galaxy formation models.

In the context of the standard, hierarchical paradigm (\lcdm), the
merger rates of distinct dark matter halos can be predicted robustly.
In particular, several studies have demonstrated that the predicted
rate of major mergers of galaxy-sized cold dark matter halos increases
with redshift as $(1+z)^m$, where the exponent lies in the range $ 2.5
\lsim m \lsim 3.5$ \citep[e.g.][]{Governato99, Gottlober01}.
Na{\"{\i}}vely, one might expect that the fraction of galaxies with
close companions (or the fraction that is morphologically identified
to be interacting) should increase with redshift according to a
similar scaling.  As we discuss below, the connection between merger
rates and the redshift dependence of close companion counts is not so
straightforward.

Observational analyses using very  different techniques to measure the
redshift evolution in the fraction  of galaxies in  pairs yield a very
broad range of evolutionary exponents, $m \simeq 0-4$
\citep[e.g.,][]{Carlberg2000,LeFevre2000,P2002,Conselice03,Bundy04}.
With the exception of the  Second Southern Sky Redshift Survey (SSRS2)
at  low  redshift  \citep{P2000},  these measurements  use  incomplete
redshift surveys  that are often  deficient in close pairs  because of
mechanical spectrograph constraints.   Thus, apparent discrepancies in
these  studies  may result  from  differing  definitions  of the  pair
fraction,  cosmic  variance, survey  size  and  selection, and  survey
completeness.

One  of the  most recent  explorations  of the  redshift evolution  of
close-companion  counts  was   performed  by  the  Deep  Extragalactic
Evolutionary  Probe   2  (DEEP2)  team   \citep{Lin04},  who  reported
surprisingly weak evolution in the pair fraction of galaxies out to $z
\sim 1.1$.   In particular, they  reported that the average  number of
companions  per galaxy,  $\Nc$,  grew with  redshift  as $N_c  \propto
(1+z)^m$,  where  $m =  0.51  \pm 0.28$.   At  face  value, this  weak
evolution seems to be in drastic  conflict with the strong, $m \sim 3$
evolution  predicted for  dark  matter halos.   Other recent  analyses
using  different data  sets and  different techniques  yield similarly
``weak'' evolution \citep[e.g.][]{Belletal06,Lotz06}.

The  resolution  of this  apparent  conflict  involves the  difference
between distinct halos and subhalos.  The ``predicted'' merger rate of
$m  \sim 3$ applies  only to  {\em distinct}  dark matter  halos.  The
prediction for {\em galaxies}, on the other hand, is more complicated.
Distinct halos are  predicted to contain $\sim 10\%$  of their mass in
self-bound   substructures  known  as   dark  matter   {\em  subhalos}
\citep[e.g.][]{klypin_etal:99}.  These subhalos  are the natural sites
for satellite galaxies around  luminous central objects.  Subhalos can
orbit  within  their  host  halo  for  a  time  that  depends  on  the
satellite-to-host mass  ratio and on specific  orbital parameters, and
can  appear  as close  companions  in  projection without  necessarily
signaling  an  imminent galaxy-galaxy  merger.   Beyond this,  precise
predictions  for the  merger rates  of galaxies  are sensitive  to the
poorly  -understood process  of  galaxy formation  inside dark  matter
halos.   In  work  that   is  based  on  dissipationless  cosmological
simulations, galaxy  formation unknowns  can be absorbed  into various
prescriptions for  assigning observable galaxies to  dark matter halos
and    subhalos   \citep[for    related    discussions   see,    e.g.,
][]{klypin_etal:99,   Bullock_etal:00,   diemand_etal04,   gao_etal04,
nagai_kravtsov:05, conroy_etal05, fd:06,wang:06, weinberg_etal:06}.

In   this   work we  use a  large   cosmological  N-body simulation to
calculate the background dark matter halo distribution and an analytic
formalism to predict the subhalo populations within each of these {\em
host} halos.  There are several noteworthy advantages to adopting this
approach.  First, we rely on straightforward mappings between galaxies
and  dark matter halos and subhalos.    Second, we measure the average
number of close companions, $N_c$,  in our simulations in exactly  the
same way in which they are observed in the real universe.  We then use
the  observed  companion counts directly,  rather  than {\em inferred}
merger rates, to discriminate between simple and physically-reasonable
scenarios for associating dark matter halos and subhalos with luminous
galaxies.  Our approach also  allows us to overcome  several technical
issues.  Using analytic models  with no inherent resolution  limits to
model substructure allows us to  overcome possible issues of numerical
overmerging in the very dense environments
\citep[e.g.,][]{klypin_etal:99}  where close  galaxy  pairs are  often
found.  Furthermore,  our analytic subhalo  model allows us  to assess
the variance in close-companion counts that can be associated with the
variation   in  substructure   populations   with  fixed   large-scale
structure.  Our  methods also allow us  to test for  the importance of
cosmic variance and  chance projections.  We find these  effects to be
small for typical close companion  criteria at redshifts less than $z
\simeq 2$ in our preferred model.

A powerful way to quantify and parameterize the way in which close
pairs constrain the relationship between halos and galaxies is through
the halo occupation distribution (HOD) of galaxies within dark matter
halos.  The HOD is the probability that a distinct, host halo of mass
$M$ contains $N$ {\em observed} galaxies, $P(N|M)$, and is usually
parameterized in a simple, yet physically-motivated way (e.g.,
\citealt{berlind_weinberg:01}, \citealt{Kravtsov04a} and references
therein).  Coupling the HOD with a prescription for the spatial
distribution of galaxies within their host halos allows for an
approximate, analytic calculation of close pair statistics
\citep*{bws:02}.  Using pair counts to constrain the HOD and the
distribution of galaxies within halos provides a direct means to
separate secure cosmological predictions for dark matter halo counts
from uncertainties in the more complicated physics operating on the
scales of individual halos and subhalos (e.g., star formation triggers
and orbital evolution).  As we discuss in \S~4, the HOD methodology
also provides a useful physical platform for interpreting our specific
simulation results and for extending them to provide a general
constraint on galaxy formation models.

The outline of this paper  is as follows.  In \S~\ref{sec:methods}, we
outline  our methods,  discuss  our simulations,  and investigate  the
importance  of interlopers  in  common estimators  of  the close  pair
fraction.   We  present our  predictions  for  the companion  fraction,
$\Nc$,  in  \S~\ref{sec:results}  and  investigate its  dependence  on
galaxy  number density  and redshift.   We compare  our  general model
predictions to the observed evolution in $\Nc$ with redshift in \S~3.1
and with our own analysis of pair counts from the UZC redshift survey
and SSRS2  at $z=0$  in \S~3.2.  We discuss the importance
of cosmic variance in pair analyses in \S 3.3 
and provide general predictions
for future surveys in \S~3.4.
In  \S~\ref{sec:interp} we  present a
detailed discussion of  our results in the context  of the galaxy 
HOD,  and explain the predicted  evolution in pair
counts as a consequence of the  convolution of a halo merger rate that
increases with redshift  and a halo mass function  that decreases with
redshift.  We  conclude in  \S~\ref{sec:discussion} with a  summary of
our primary results and a discussion of future directions.

\section{Methods}
\label{sec:methods}

The following subsections detail our theoretical methods.  Briefly, we
investigate pair count statistics using an $N$-body simulation to
account for large-scale structure and the host dark matter halo
population as described in \S~\ref{sec:simulations}.  We use an
analytic substructure model \citep{Zentner05} to identify satellite
galaxies within these host halos as described in
\S~\ref{sec:submodel}.  This ``hybrid'' technique overcomes
difficulties associated with numerical overmerging on small scales in
the large cosmological simulation box and has already been
demonstrated to model accurately the two-point clustering statistics
of halos and subhalos \citep{Zentner05}.  We use a simple method to
embed galaxies in our simulation volume (\S~\ref{sec:model}) and
``observe'' these mock galaxy catalogs in a manner similar to those
used in observational studies.  We then compute pair fraction
statistics that mimic those applied to observational samples
(\ref{sec:nc}).

%
%
%
\begin{figure}[t!]
\epsscale{1.05}
\plotone{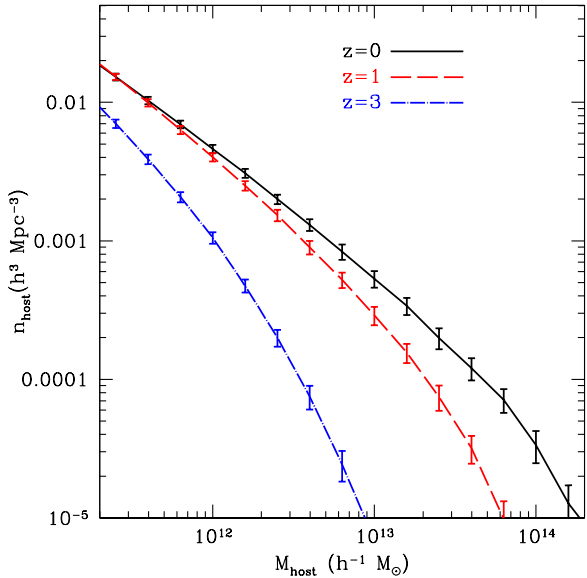}
\caption{ The cumulative mass function of host halos as derived from
  our $120$~$\hMpc$ simulation box at $z = 0$,~$1$, and $3$.  Error
  bars estimate cosmic variance using jackknife errors from the eight
  octants of the computational volume.}
\label{fig:massfunc}
\end{figure}
%
%
%

\subsection{Numerical Simulation} 
\label{sec:simulations}
Our  simulation was  performed using  the
Adaptive Refinement Tree  (ART) $N$-body code \citep{kravtsov_etal:97}
for  a flat  Universe with  $\Omega_m=  1 -  \Omega_{\Lambda} =  0.3$,
$h=0.7$, and $\sigma_8 =  0.9$.  The simulation followed the evolution
of $512^3$ particles in comoving box  of $120 \hmpc$ on a side and has
been previously discussed in \citet{Allgood05} and \citet{Wechsler05}.
The  corresponding  particle mass  is  $m_p  \simeq  1.07 \times  10^9
\hMsun$.  The  root computational grid was comprised  of $512^3$ cells
and  was adaptively refined  according to  the evolving  local density
field  to a  maximum  of $8$  levels.   This results  in peak  spatial
resolution of $h_{\mathrm{peak}} \simeq 1.8 \hkpc$ in comoving units.

We identify  host halos  using a variant  of the Bound  Density Maxima
algorithm \citep[BDM,][]{klypin_etal:99}. Each halo is associated with
a density  peak, identified  using the density  field smoothed  with a
24-particle  SPH  kernel  \citep[see][for  details]{Kravtsov04a}.   We
define  a halo virial  radius $\Rvir$,  as the  radius of  the sphere,
centered  on  the density  peak,  within  which  the mean  density  is
$\Dvir(z)$ times  the mean density  of the universe,  $\rhomean$.  The
virial  overdensity  $\Dvir(z)$, is  given  by  the spherical  top-hat
collapse approximation and we compute it using the fitting function of
\citet{bryan_norman:98}. In  the $\Lambda$CDM cosmology  that we adopt
for our simulations, $\Dvir(z=0) \simeq 337$ and $\Dvir(z) \rightarrow
178$  at  $z  \gsim 1$.   In  what  follows,  we  use virial  mass  to
characterize the masses of distinct host halos (halos whose centers do
not lie within  the virial radius of a larger  system).  The host halo
catalogs  are complete  for virial  masses $M  \gsim  10^{11} \hMsun$.
Figure \ref{fig:massfunc} shows the host halo mass functions at $z=0$,
$1$,  and $3$  resulting  from  this procedure.   The  error bars  are
jackknife  errors computed from  the eight  octants of  the simulation
volume and  reflect uncertainty  in the halo  abundance due  to cosmic
variance.

%
%
%
\begin{figure}[t!]
\epsscale{1.3}
\plotone{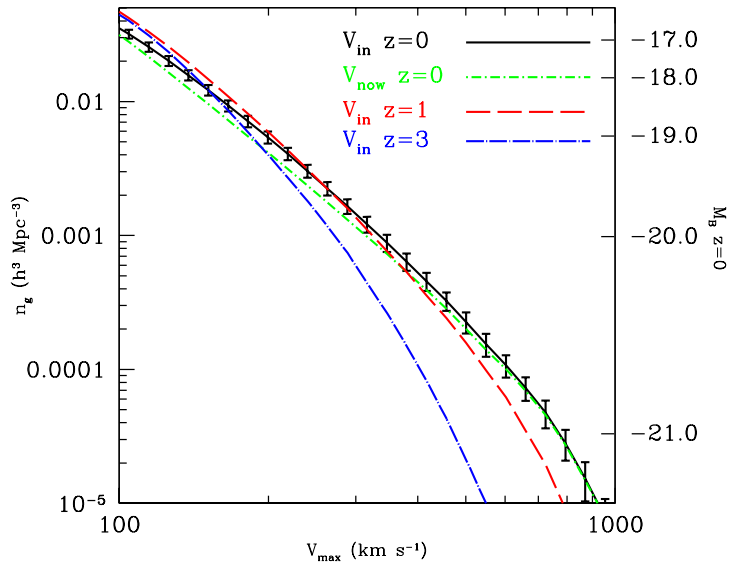}
\caption
{ 
The  left-hand vertical axis  shows the  cumulative number  density of
galaxies in  our simulation  catalog as a  function of  velocity using
$V_{now}$ at  $z=0$ (dot-dash, current $\Vmax$) and  $V_{in}$ at $z=0$
(solid, the value  of $\Vmax$ when the halo  was accreted) to identify
subhalos  as galaxies.   The  right-hand axis  maps  the $z=0$  B-band
absolute magnitude  of galaxies from the SSRS2  luminosity function to
the appropriate galaxy number  density shown on the left-hand vertical
axis.   Using the  right-hand scale,  the solid  and  dot-dashed lines
demonstrate our  adopted $z=0$ relationship between  $\Vin$ or $\Vnow$
and  galaxy luminosity.   The remaining  two lines  show  the $V_{in}$
sample  at  $z=1$ and  $z=3$.   A  comparison  with Figure  1  reveals
dramatic  differences  between the  evolution  of  distinct host  halo
number  densities  and the  galaxy  (subhalo)  number densities  shown
here. The errorbars shown were  generated by summing in quadrature the
jackknife error and the realization to realization scatter.
}
\label{fig:vmf}
\end{figure}
%
%
%

\subsection{Substructure Model} 
\label{sec:submodel}

In order to determine the substructure properties in each host dark
matter halo, we model halo mass accretion histories and track their
substructure content using an analytic technique that incorporates
simplifying approximations and empirical relations determined via
numerical simulation.  We provide a brief overview of the technique.
The model is described in detail in \citet[][hereafter
  Z05]{Zentner05}, and is an updated and improved version of the model
presented in \citet{zb:03}.  This model shares many elements with
similar models of \citet{taylor_babul04}, \citet{penarrubia_benson05},
and \citet{vandenbosch_etal05}.

In hierarchical CDM-like models, halos accumulate their masses through
a series of mergers with smaller objects.  Thus the first step of any
analytic calculation is to build an approximate account of this merger
hierarchy.  For each host halo of mass $M$ at redshift $z$ in our
numerical simulation volume, we randomly generate a mass accretion
history using the extended Press-Schechter formalism
\citep{Bond91,LC93} with the particular implementation of
\citet{Somerville99}.  The merger tree consists of a list of all of
the distinct halo mergers that have occurred during the process of
accumulating the mass of the final host object at the redshift of
observation.  Every time there is a merger, the smaller object becomes
a subhalo of the larger object.

We then  track the evolution  of subhalos as  they evolve in  the main
system  in the  following way.   At each  merger event,  we  assign an
initial orbital  energy and angular  momentum to the  infalling object
according  to the probability  distributions culled  from cosmological
$N$-body  simulations by  Z05.  We  then  integrate the  orbit of  the
subhalo in the  potential of the main halo from  the time of accretion
to the epoch of observation.  We model tidal mass loss with a modified
tidal approximation  and dynamical friction  using a modified  form of
the   Chandrasekhar  formula   \citep{chandrasekhar43}   suggested  by
\citet{hashimoto_etal03}.   For  simplicity,   we  model  the  density
profiles  of   halos  using  the   \citet[][NFW]{NFW97}  profile  with
concentrations  set  by their  accretion  histories  according to  the
algorithm  of  \citet{wechsler:02}.  As  subhalos  orbit within  their
hosts, they  lose mass and their maximum  circular velocities decrease
as  the  profiles  are  heated  by tidal  interactions.   A  practical
addition  to this  evolution  algorithm is  a  criterion for  removing
subhalos from catalogs  once their bound masses become  so small as to
render them  very unlikely  to host a  luminous galaxy.   This measure
prevents  using  most of  the  computing  time  to evolve  very  small
subhalos on very tightly bound  orbits with very small timesteps.  For
the purposes  of the present work,  we track all  subhalos until their
maximum  circular velocities drop  below $\Vmax  = 80$~\kms,  at which
point they are  no longer considered and are  removed from the subhalo
catalogs.  We refer the interested reader  to \S~3 of Z05 for the full
details of this model.

The computational demands of this  analytic model are such that we can
repeat this process several times for each host halo in the simulation
volume in  order to determine the  effect of the  variation in subhalo
populations among halos of fixed  mass on our close pair results.  For
example, one  source of  noise is due  to subhalo orbits:  subhalos at
pericenter are likely  to be counted as close  pairs, whereas subhalos
at apocenter are  less likely to be counted as  close pairs.  In order
to  make  this assessment,  we  repeat  this  procedure for  computing
subhalo  populations for each  host halo  in the  computational volume
thereby creating distinct  subhalo catalogs with identical large-scale
structure  set by  the large-scale  structure of  the  simulation.  We
refer to each of these  distinct subhalo catalogs as a ``realization''
of the model, a term  which derives from the inherent stochasticity of
the  model. Additionally we  may perform  rotations of  the simulation
volume.   These rotations  provide us  with different  lines  of sight
through the  substructure of  the simulation. Since  the perpendicular
separation  along the  line  of  sight of  the  subhalos is  extremely
important  to  the close  pair  statistics,  along  with the  velocity
differences  along these lines  of sight,  these rotations  provide us
with  additional  effective  ``realizations''.   Three  rotations  are
performed  on each simulation  box, providing  us with  more effective
``realizations''.

This   substructure  model   has  proven   remarkably   successful  at
reproducing  subhalo  count   statistics,  radial  distributions,  and
two-point  clustering  statistics  measured in  full,  high-resolution
$N$-body  simulations   in  regimes  where  the   two  techniques  are
commensurable.   The results of  the model  agree with  full numerical
treatments over  more than $3$ orders  of magnitude in  host halo mass
and as a  function of redshift (Z05).  This  success motivates its use
to overcome  some of the  difficulties associated with using  a purely
numerical  treatment.  Specifically,  unlike N-body  simulations, this
method  suffers  from no  inherent  resolution  limits  and makes  the
process of selecting subhalo populations based on specific features of
their merger  histories or detailed orbital evolution  clean and easy.
Note that any correlation between formation history or the HOD and the
larger scale environment will not be included in catalogs created with
this hybrid method, but this is  not expected to be important for most
of the mass range we consider here \citep{Wechsler05}.

We  explore  two simple  yet  reasonable  toy  models for  associating
galaxies with dark matter subhalos.   In order to quantify the size of
subhalos, we use the  subhalo maximum circular velocity, $\Vmax \equiv
max[  \sqrt{G M(<r)/  r}  ]$.   This choice  is  motivated by  several
considerations.   As long  as only  systems  that are  well above  any
resolution  limits  are considered,  this  quantity  is measured  more
robustly in simulation  data and is not subject  to the same ambiguity
as particular  mass definitions because $\Vmax$  is typically achieved
well  within the  tidal radius  of a  subhalo.  These  facts  make our
enumerations  by  $\Vmax$ easier  to  compare  to  the work  of  other
researchers using different subhalo identification algorithms.  In the
next section  we describe our  mapping of galaxies onto  subhalos, and
explore models for mapping galaxies onto subhalos that use the maximum
circular velocities defined at two different epochs.

%
%
\begin{figure*}[t]
\epsscale{1.0}
\plottwo{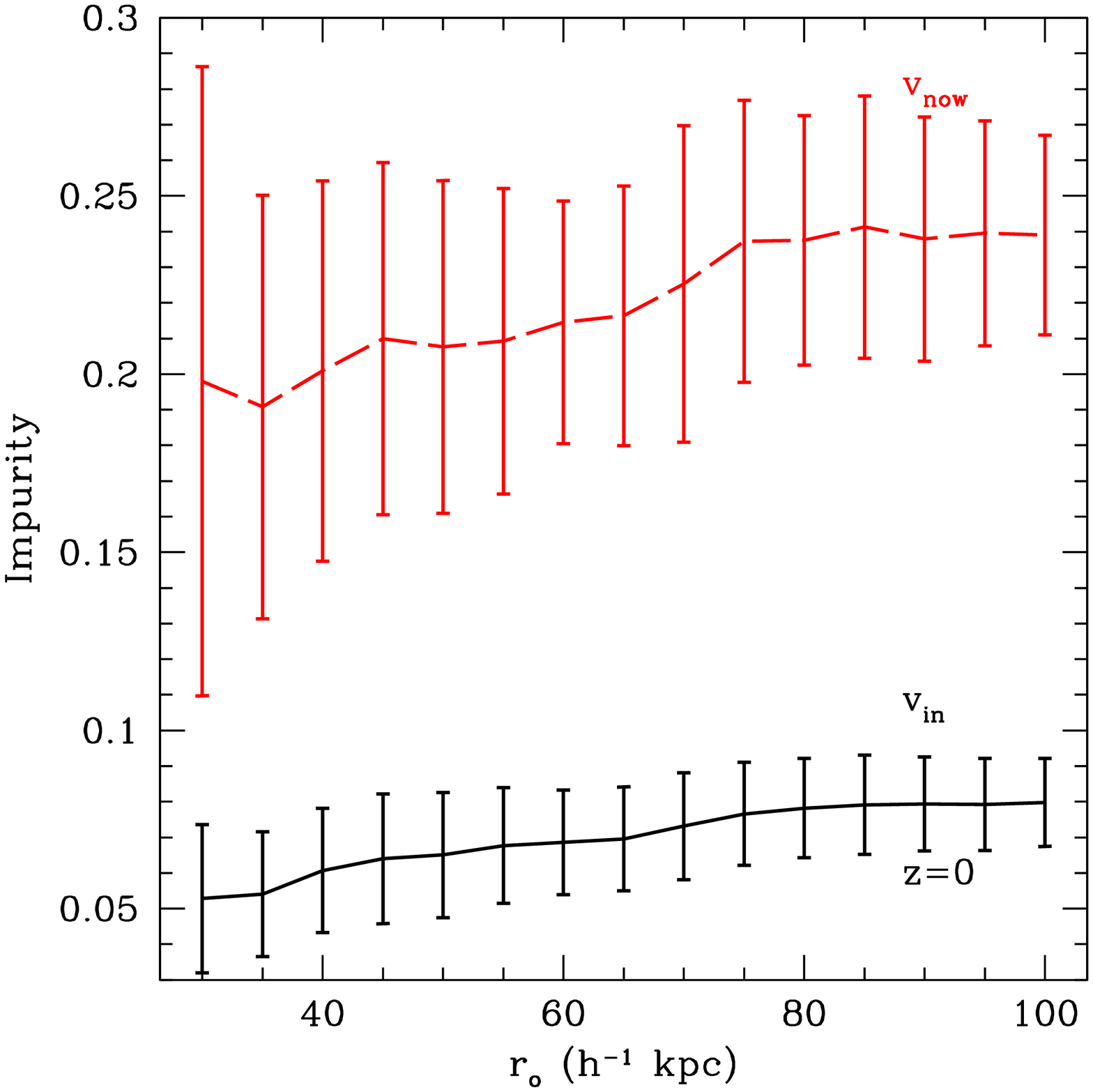}{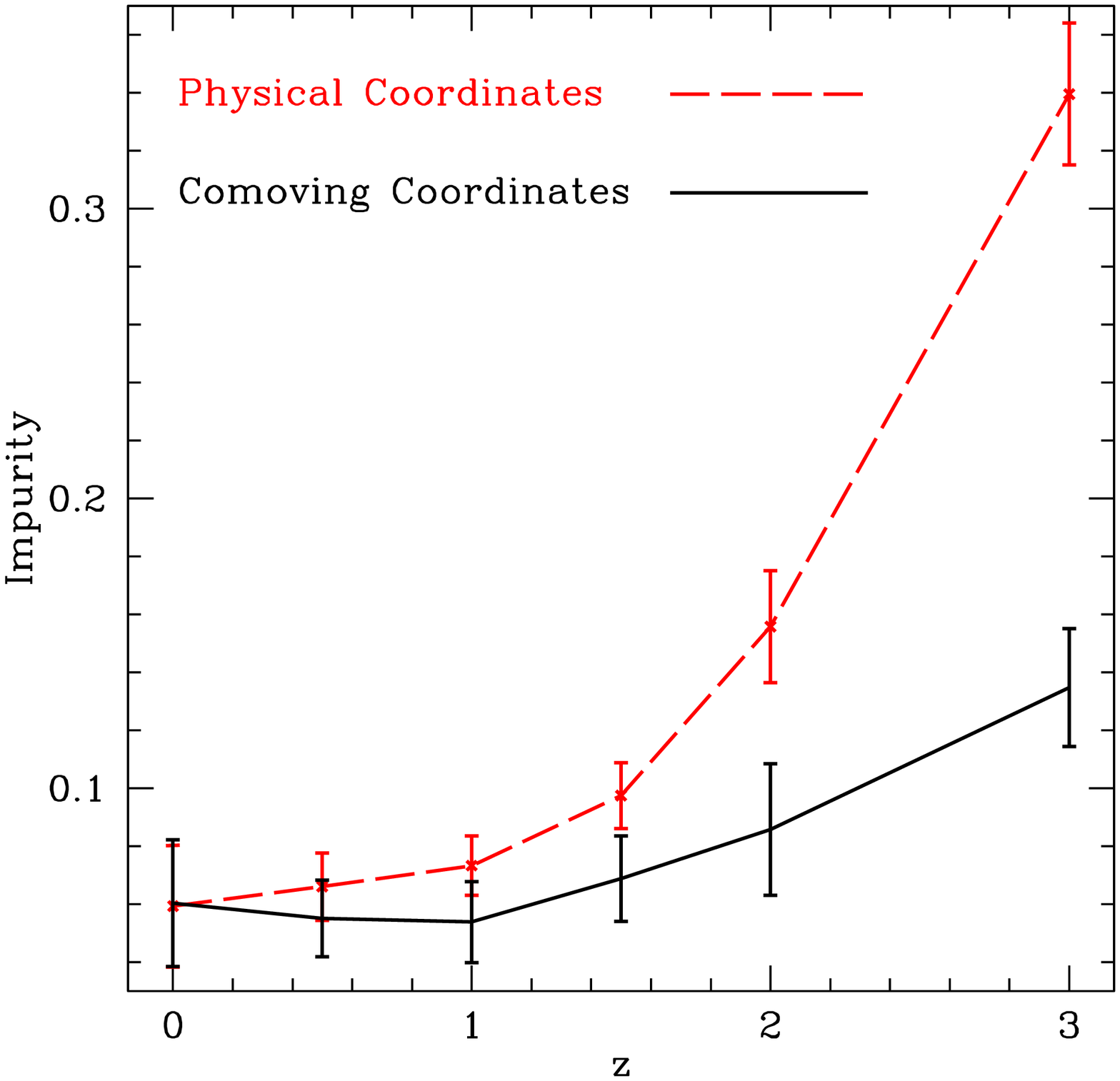}
\caption{
Pair  count impurity  in our  simulated galaxy  catalogs with  a fixed
comoving number  density $n_g =  0.01 h^{3}$ Mpc$^{-3}$.   Impurity is
defined to be the fraction of pair-identified galaxies that do not sit
within  the same  host dark  matter  halo.  The  error bars  represent
variations in the impurity due to cosmic variance as well as variation
in the  subhalo populations.  The uncertainties  were calculated using
the realization to realization scatter and the jackknife errors.  {\em
Left:} The $z=0$ impurity as a function of outer projected radius used
for the  pair criteria ($r_o$  in the text).   Here we have  fixed the
inner radius for pair identification at $r_i = 10 ~\hkpc$ and demanded
line-of-sight velocity  separations to be  within $\Delta v =  \pm 500
$\kms. The solid line represents  the $V_{in}$ sample while the dashed
line  uses $V_{now}$.   {\em Right:}  The  impurity as  a function  of
redshift within  a fixed physical separation of  $r_o =  50$ h$^{-1}$
kpc, using $V_{in}$.  The red dashed  line is the impurity in the pair
fraction for a  system in physical coordinates.  The  black solid line
is the impurity  in the comoving coordinates. Note  that in most cases
$\gsim  90  \%$  of  galaxies  identified  as  close  pairs  by  these
observationally-motivated criteria  inhabit the same  host dark matter
halo for $z \leq 1.5$. In the physical coordinates we see a sharp rise
in the impurity beyond this redshift.
}
\label{fig:Purity}
\end{figure*}
%
%
%

\subsection{Assigning Galaxies to Halos and Subhalos} 
\label{sec:model}

After  computing the  properties  of  halos and  subhalos  in a  \lcdm
cosmology, the  next step is to  map galaxies onto  these objects.  In
our comparisons with observational data, we normalize our model galaxy
catalogs  to an  observed  population by  matching  their mean  number
densities to  those of  halos and subhalos.   In each case,  we assume
that there is a monotonic relationship between halo circular velocity,
$\Vmax$, and galaxy luminosity in  the relevant band of interest.  The
number density of galaxies as  a function of halo (subhalo) $\Vmax$ is
shown in Figure \ref{fig:vmf}.  For our $z=0$ comparisons we adopt the
SSRS2  B-band  luminosity  function  in  order to  match  the  samples
appropriately.  The  implied mapping  between $M_{\rm B}$  and $\Vmax$
for the SSRS2 luminosity function  is demonstrated in the right y-axis
of  Figure \ref{fig:vmf}.   For the  \cite{Lin04} DEEP2  comparison at
$z\simeq 0.5 -  1.1$ we match our $\Vmax$  functions with the measured
luminosity function in the $R_{\rm AB}$ band.

Although a monotonic relationship  between $\Vmax$ and luminosity is a
simplifying assumption, it serves as  a useful, physical model for our
comparisons, and  is motivated in part by  the well-known Tully-Fisher
relation which shows that the  {\it observed} speed of a spiral galaxy
correlates with its luminosity  in all bands \citep{TF77}.  Scatter in
this relationship could  (of course) be included in  our model, but we
have chosen to  neglect this for the sake  of simplicity.  Moreover, a
number of studies have demonstrated that threshold samples selected in
this  way exhibit  excellent  agreement with  many  of the  clustering
statistics of observed  galaxy samples \citep[e.g.,][; Marin, Wechsler
\&       Nichol,      in      preparation]{klypin_etal:99,Kravtsov04a,
tasitsiomi_etal04,  azzaro_etal05,  conroy_etal05}.   The decision  to
normalize  our catalogs  based on  the number  densities of  halos and
subhalos above  thresholds in maximum  circular velocities circumvents
the  need to  model star  formation  directly in  our simulations  and
allows  us to  focus on  well-understood physical  parameters.   As we
demonstrate  below, close  pair  fractions vary  strongly with  galaxy
number density at fixed redshift.

For subhalos,  a monotonic relationship between  {\em current} maximum
circular velocity and galaxy luminosity is less well-motivated than it
is for  host halos (or  halos in the  ``field'').  When a  halo merges
into a  larger host  halo and becomes  a subhalo,  it loses mass  as a
result  of  tidal  interactions  and  its  maximum  circular  velocity
decreases.   The  degree  to  which  luminosity  is  lost  or  surface
brightness  declines   is  considerably  more   uncertain  \citep[see,
e.g.][and Bullock \& Johnston,  in preparation]{bj05}, but is probably
less than the decrease in mass \citep[e.g.][]{nagai_kravtsov:05} Using
maximum  circular velocities  rather  than total  bound masses  partly
accounts for this mismatch between mass and luminosity because maximum
circular  velocity scales  only very  slowly with  bound  mass, $\Vmax
\propto      M^{\gamma}$,       with      $\gamma      \sim      0.25$
\citep{Kravtsov04b,kazantzidis_etal04}.   In  addition,  environmental
effects  add  a  further   complication  as  they  may  influence  the
luminosity or star formation rate of a galaxy in such a way that it is
driven off  of any $\Vmax$-luminosity  relationship that may  exist in
the field.

We  deal with  this  uncertainty  by adopting  two  simple models  for
associating  subhalos  with  galaxies:  first,  the  maximum  circular
velocity that the subhalo had when it was first accreted into the host
halo,  $\Vin$, and  second,  the maximum  circular  velocity that  the
subhalo  has at  the current  epoch,  $\Vnow$.  The  choice of  $\Vin$
mimics a case  where a galaxy is much more  resistant to reductions in
luminosity than the  dark matter subhalo is to  mass loss.  The choice
of $\Vnow$ represents a case where  a galaxy drops out of a luminosity
threshold sample (or, perhaps more realistically, a surface-brightness
threshold sample) in  direct proportion to the decline  in the maximum
circular velocity of the subhalo.

As we  stated above,  subhalo $\Vmax$ declines  much more  slowly than
total bound mass during episodes  of mass loss, so both methods assume
that  the luminosity  of  the  galaxy is  resistant  to large  changes
resulting from the early stages  of interaction, which is supported by
observations  of pairs  at low  redshifts \citep{Barton01a,Barton:03}.
In the case of $\Vin$, the luminosity is set in the field and does not
change  after merging  into  the host  system,  while in  the case  of
$\Vnow$  the luminosity declines  slowly as  a result  of interactions
with the host system.

The $\Vin$ choice  is likely the most natural one,  and has been shown
to provide an excellent match to clustering statistics over a range of
redshifts  for scales larger  than 100  $\hkpc$ \citep{conroy_etal05}.
However, the  $\Vnow$ model provides  acceptable clustering statistics
at large  separations \citep{Kravtsov04a},  and provides a  very useful
reference   point  for  exploring   the  uncertainty   in  theoretical
predictions of close  pair fractions and for highlighting  the type of
physical processes that close pairs can and {\em cannot} probe.

We emphasize that the evolution of the host halo merger rate is not an
appropriate comparison  for the evolution of close  pair counts, which
probe  the galaxy  or  subhalo  merger rate.   The  models we  examine
represent two different choices for how galaxies populate their hosts.
The fact  that the $\Vin$ and  $\Vnow$ models both have  the same host
halo merger rates, but different halo occupation counts that result in
very different predictions  for small-scale clustering, highlights the
fact that close pairs are  not directly connected to halo merger rates
except in  the context of a  specific model for  how galaxies populate
these halos.

Figure   \ref{fig:vmf}  shows   the  cumulative   number   density  of
``galaxies'' identified  in our simulations,  $n_g$, as a  function of
their  maximum  circular  velocities.   The  solid  line  shows  $z=0$
galaxies using $\Vin$  as an identifier.  The dash-dot  line shows the
same  using $\Vnow$.   The $\Vnow$  choice lies  below  $\Vin$ because
galaxies more  readily fall  out of the  observational sample  in this
case due to  the mass lost by the subhalo  after accretion.  Note that
host halos always have $\Vnow =  \Vin$.  The fact that the $\Vnow$ and
$\Vin$ number  density lines differ  by only $\sim 30\%$  reflects the
fact that  most galaxies  are field galaxies  and this  choice affects
only the  satellites of group  and cluster systems.  The  right y-axis
for the  solid and dash-dot lines in  Figure \ref{fig:vmf} illustrates
how  this choice  affects  our mappings  between  velocity and  galaxy
luminosity at  $z=0$.  The dashed  and long-dash-dot lines  show $n_g$
for the  $\Vin$ choice  at $z=1$ and  $z=3$.  The  luminosity function
used  is the combined  SSRS2 luminosity  function \citep{Marzkeeta98}.
Compared with  Figure \ref{fig:massfunc} we see that  the {\em galaxy}
velocity function evolves much more weakly with redshift than does the
{\em host halo} mass function.

\subsection{Defining Close Galaxy Pairs and 
Diagnosing the Effect of Interlopers} 
\label{sec:nc}

Various  definitions have  been used  for identifying  close  pairs of
galaxies  \citep[e.g.,][]{Kennicutt:84,  z&k89, Burkey94,  Carlberg94,
Woods95, Y&E95, Barton:00,  P2000, Carlberg2000, Bundy04, LeFevre2000,
P2002,P97, Neus97}.   In this  work we follow  the definition  used in
\citet{P2002}  and  \citet{Lin04}, and use the  average  number  of
companions per galaxy to enumerate close galaxy pairs:
\begin{equation}
N_c \equiv \frac{2 n_{\rm p}}{n_g}.
\label{eqt:nc}
\end{equation}
In Eq.~(\ref{eqt:nc}), $n_g$ is the  number density of galaxies in the
sample and $n_{\rm p}$ is the number density of individual pairs.

We  define   pairs  as  galaxies  that  fall   within  a  well-defined
line-of-site velocity separation $\vert  \Delta v \vert$ and that have
a projected, center-to-center separation on the sky, $r_{\rm p}$, that
lie between  an inner and  outer value $r_i  < r_{\rm p} <  r_o$.  The
inner radius is chosen  to prevent morphological confusion.  The outer
radius  may be  adjusted  to  reduce contamination  of  the sample  by
interlopers.  We use fiducial values of $r_i = 10 \hkpc$ and $r_o = 50
\hkpc$   in  physical  units   and  we   adopt  a   fiducial  relative
line-of-sight  velocity   difference  of  $\vert  \Delta   v  \vert  =
500$~\kms,  following   \citet{P2002}  and  \citet{Lin04}.    In  what
follows, we refer  to the chosen close-pair volume  as a ``cylinder''.
We  emphasize that  in our  standard measure  we adopt  {\em physical}
radii and  a constant  $\Delta v$  cut to define  our cylinder  at all
redshifts (``physical  cylinder''), but we  also explore a  case where
the radii and  line-of-sight depth (defined by $\Delta  v$) are scaled
with the expansion of the universe (``comoving cylinder'').

Our model catalogs provide a useful tool for studying the degree of
contamination by interlopers in this commonly-adopted measure.  If we
define interlopers as projected galaxy pairs that {\em do not} lie
within the same host halo, we can define a measure of close pair count
{\em impurity} as
\begin{equation}
\label{eq:Imp}
{\rm Impurity} \equiv \frac{n_f}{n_p},
\end{equation}
where $n_f$  is the number density  of interloping, or  false pairs in
the sample and $n_p$ is the total number density of observed pairs.

The  left  panel of  Figure~\ref{fig:Purity}  shows  how the  impurity
varies as a function of the choice of outer pair radius $r_o$ for both
$V_{in}$ (solid)  and $V_{now}$ (dashed)  samples at $z=0$.   For each
choice, we fix $r_i = 10 \hkpc$ and define the galaxy population using
a  galaxy number  density $n_g=  0.01 h^{3}$~Mpc$^{-3}$.   The $\Vnow$
case has  a higher  overall impurity ($\sim  20 \%$ compared  to $\sim
5\%$), reflecting  the fact that  the amplitude of $\Nc$  is generally
lower for $\Vnow$ (see \S 3 ) and is therefore more affected by chance
projections.  Both  the $\Vin$  and $\Vnow$ samples  show only  a mild
increase  ($\sim  3 \%$)  in  the  interloper  fraction as  the  outer
cylinder radius is varied from $r_o = 30$ $\hkpc$ to $100$ $\hkpc$.

The  right  panel of  Figure~\ref{fig:Purity}  shows  the impurity  as
function of redshift for  the $\Vin$-selected sample.  The dashed line
corresponds to a physical cylinder and the solid line corresponds to a
comoving cylinder at the same fixed comoving number density $n_g= 0.01
h^{3}$~Mpc$^{-3}$.  Both lines have $r_o = 50 \hkpc$.  The impurity in
the comoving cylinder remains relatively small with $z$, while for the
physical  cylinder, the  fraction of  interlopers remains  small until
$z\sim 2$, after which it  rises sharply.  This reflects the fact that
halo  virial radii  scale as  $\sim (1+z)^{-1}$  and become  a smaller
fraction of a fixed physical  $r_o$ at high redshift.  While one might
interpret this result  as favoring the use of  a comoving cylinder for
pair  identification, there  are practical  difficulties  in resolving
pairs at  small comoving separations  at high redshift.   For example,
while physical separations between $r_p  = 10 - 50$ $\hkpc$ correspond
to resolvable angular  separations at $z = 2$,  $\theta_p \simeq 1.7 -
8.5$ arc  seconds, the corresponding comoving distances  would be much
more difficult  to resolve, $\theta_p  \simeq 0.57 - 2.8$  arc seconds.
Ground   based  surveys   require  separations   $>  2$   arc  seconds
\citep{Belletal06}.  The  main conclusion to draw from  this figure is
that  for  most  practical  pair  measures,  interlopers  will  be  of
increasing importance at high redshift.

%
%
%
\begin{figure}[t]
\epsscale{1.0}
\plotone{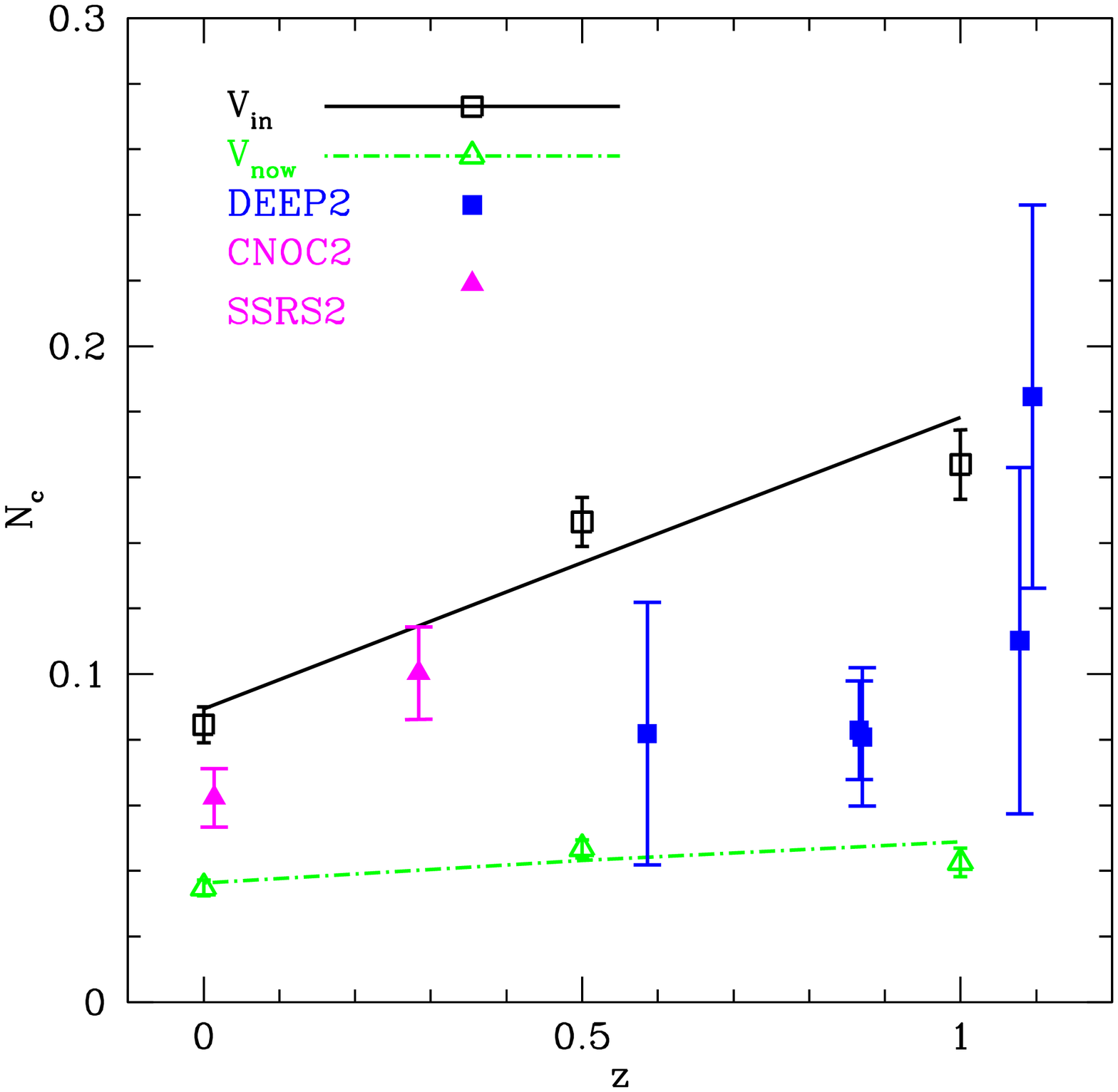}
\caption{
Evolution in  the companion fraction.   Solid squares are  data points
from the DEEP2 Survey taken  from fields 1 and 4, \citep{Lin04}. Close
companions are defined as  having physical separations between $r_p =
10  - 50$  $\hkpc$ with  a  line of  sight velocity  difference of  $<
500$~\kms.    Solid  triangles   are  data   from  CNOC2   and  SSRS2,
\citep{P2002}.  The  empty points with  best-fit lines are  taken from
our  simulations.   At  each  redshift  we select  model  galaxies  by
approximately matching  the number densities  of galaxies used  in the
DEEP2,  SSRS2, and CNOC2  points to  the nearest  simulation redshift.
Empty  triangles  use  $\Vnow$   velocity  function  to  match  number
densities while empty squares use  $\Vin$.  Lines show the best-fit to
$\Nc \propto (1+z)^m$  with $m=0.42 \pm 0.17$ ($\Vnow$)  and $m = 0.99
\pm 0.14$  ($\Vin$).  In both  cases the predicted evolution  is quite
weak and broadly consistent with the data.
}
\label{fig:DEEP2_compare}
\end{figure}
%
%
%
%
\begin{figure*}[t!]
\epsscale{1.0}
\plotone{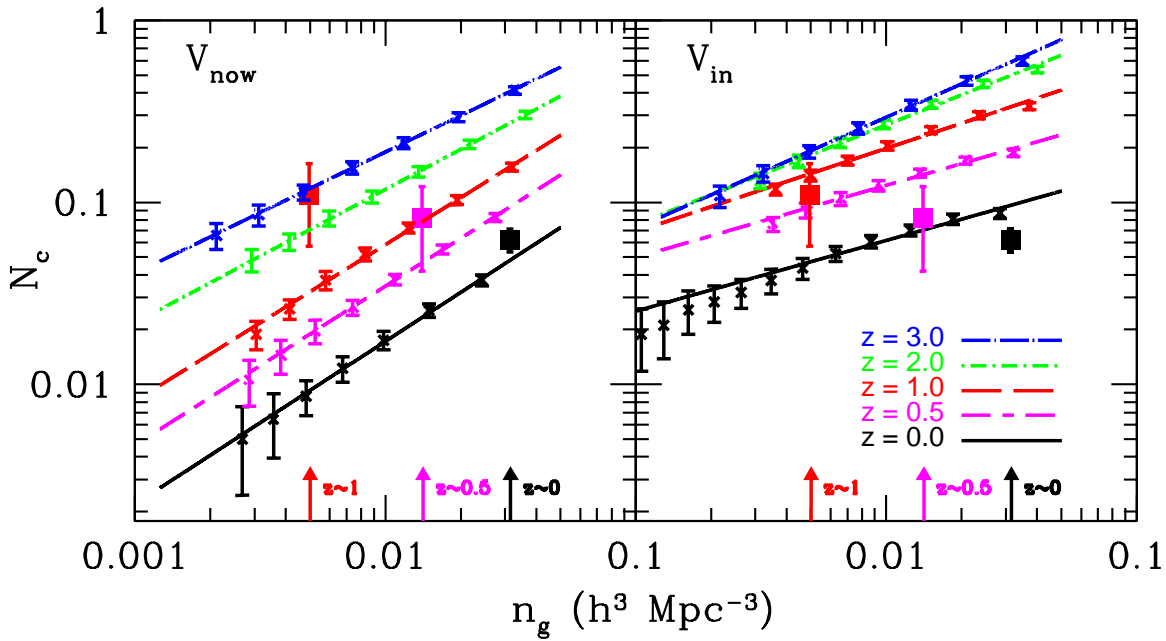}
\caption{
Close companion fraction as a  function of number density and redshift
for a  fixed {\em  physical} separation between  $r_i = 10  \hkpc$ and
$r_o = 50 \hkpc$ in projection and a line-of-sight velocity difference
$\vert \Delta  v \vert \le  500$~\kms.  Small points with  linear fits
show the the  average number of close companions  $\Nc$, as a function
of galaxy  number density $n_g$,  for our model with  galaxy selection
based  on $\Vnow$  (left panel)  and $\Vin$  (right panel).   The five
lines show power-law  fits of the companion fraction  at $z=0$, $0.5$,
$1$,  $2$, and  $3$ (from  bottom  to top)  to the  form $\Nc  \propto
n_g^a$.  The best-fit slopes vary from $a = 0.89 \pm 0.09$ at $z=0$ to
$a =  0.67 \pm 0.07$ at  $z=3$ for galaxy samples  selected on $\Vnow$
and from $a=0.39 \pm 0.07$ at  $z=0$ to $a=0.61 \pm 0.04$ at $z=3$ for
samples  selected  on  thresholds  in  $\Vin$.   Large,  solid  points
represent      observational       data      points      shown      in
Figure~\ref{fig:DEEP2_compare}.  These points are taken from the three
nearest redshift bins  to $z \simeq 0$, $0.5$,  and $1.0$, and plotted
at the number densities of  the data samples.  Arrows along the bottom
edge  of the  figure indicate  the redshift  corresponding to  each of
these data  points.  The  decreasing number densities  with increasing
redshift  in the observational  samples results  in a  nearly constant
value  of  the  companion  fraction  with  redshift.   The  data  seem
marginally to favor the $\Vin$ model over $\Vnow$.
}
\label{fig:Nc_n}
\end{figure*}
%
%

\section{Results} 
\label{sec:results}

\subsection{Evolution in Companion Counts}
\label{sub:evolution}
We first present  results on the overall evolution  of close companion
counts with redshift.   Figure~\ref{fig:DEEP2_compare} shows our model
results for the  companion fraction of galaxies, $N_c$,  as a function
of redshift  for our $\Vin$  selection (open squares, with  line fits)
and our  $\Vnow$ selection criterion (open triangles,  with line fits)
compared with  some recent observational  measurements.  Solid squares
are  data from  the \citet{Lin04}  DEEP2  analysis (fields  1 and  4).
Solid  triangles  at  lower  redshift  reflect data  from  the  Second
Canadian Network  for Observational Cosmology (CNOC2)  survey and data
from SSRS2 \citep{P2002}.

As  was  found  in  the  observational sample  of  \citet{Lin04},  the
companion  count exhibits  very weak  evolution when  plotted  in this
manner.  It  is also evident that  both the $\Vin$  and $\Vnow$ models
show similarly weak  evolution and seem to be  broadly consistent with
the  observational data.  The  formal fit  by \citet{Lin04}  from this
data gives  close pair evolution as  $\Nc \propto (1+z)^m$,  with $m =
0.51 \pm  0.28$.  A similar  fit to our  model results, from  $z=0$ to
$z=1$,  shown  in  Figure~\ref{fig:DEEP2_compare} yields  $m=0.42  \pm
0.17$ for  selection on $\Vnow$,  and $m =  0.99 \pm 0.14$  for sample
selection  based   on  $\Vin$.   Note  that  the   $\Vin$  points  are
systematically higher than the $\Vnow$ points.  This reflects the fact
that galaxies  are more easily driven  out of an  observable sample by
tidal mass  loss in the $\Vnow$-selected  sample (see \S~4  for a more
detailed discussion).

While  Figure~\ref{fig:DEEP2_compare}  seems  relatively  simple  upon
first examination, the broad  agreement between simulation results and
the data  is attained  because we have  been careful to  normalize the
data  and theory in  a consistent  manner.  Each  of the  DEEP2 points
correspond  to  {\em different}  underlying  galaxy number  densities.
This was done in order  to account for gross luminosity evolution with
redshift,  $z$, assuming  a simple  model where  $M(z) =  M(0)-z$.  In
order to make a fair comparison with their results, we have forced the
mean number density  of the model galaxies to  match the corresponding
observed number  density of galaxies  by varying the  maximum circular
velocity threshold  (using either $\Vin$ or  $\Vnow$).  This technique
corresponds  to  re-normalizing the  relationship  between $\Vin$  (or
$\Vnow$) and galaxy luminosity at each redshift.

Using the correct number density for comparison at each redshift is of
critical importance.  In  Figure~\ref{fig:Nc_n}, we plot the companion
fraction as a function of  comoving galaxy number density $n_g$ in our
model  catalogs (small  points  with line  fits)  using $\Vnow$  (left
panel)  and  $\Vin$ (right  panel)  for  five  redshift steps  between
$z=0.0$ and  $z=3.0$.  For  both models, at  fixed $z$,  the companion
fraction increases  with the overall  galaxy number density,  while at
{\em fixed}  comoving number density the  companion fraction increases
with  redshift.   The  number  densities  and close  pair  counts  for
selected data points in Figure~\ref{fig:DEEP2_compare} are represented
as  large,  solid squares  in  Figure~\ref{fig:Nc_n}.  An  approximate
redshift for  each of the data  points is indicated by  an arrow along
the bottom of the plot.  Comparing the lines with the solid squares in
Figure~\ref{fig:Nc_n}   illustrates  that   the  $\Nc(z)$   points  in
Figure~\ref{fig:DEEP2_compare}  show  little  evidence  for  evolution
simply  because they correspond  to lower  number densities  at higher
redshifts.  An  analytic explanation for how $\Nc$  scales with number
density  is  given  in  the  Appendix, where  we  also  present  $z=0$
correlation functions  for the $\Vmax$  models at different  values of
$n_g$.

\subsection{$z=0$ Companion Counts}
\label{sub:companion}

The trend  with $\Nc$ and  $n_g$ seen in  the simulations leads  us to
look for such a trend  in some observational samples.  To compare with
$z=0$  galaxies, we  must  balance  the need  for  a large  comparison
redshift  survey that  fairly samples  large-scale structure  with the
requirement for completeness with close galaxy pairs.  Our approach is
to use small  but nearly complete redshift surveys  and to examine the
effects of cosmic variance {\it a posteriori}.

%
%
%
\begin{figure}[t]
\epsscale{1.0}
\plotone{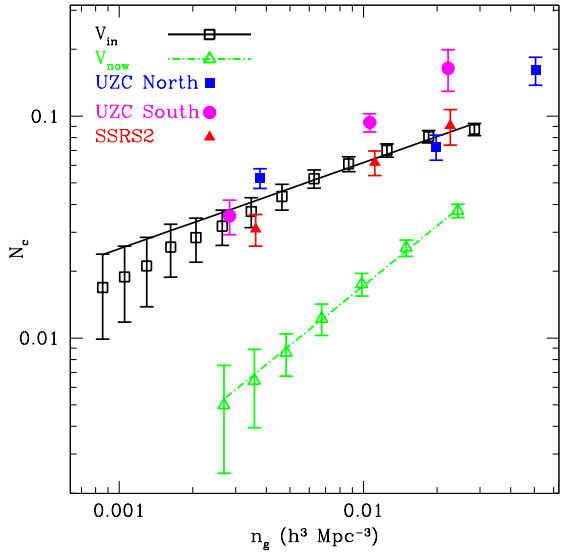}
\caption{ The variation of $N_c$ at $z=0$ based on the number density,
$n_g$, of  galaxies in the sample.   We vary galaxy  number density by
making  different threshold  cuts  in $V_{in}$  and $V_{now}$.   Close
companions are  defined as having physical separations  between $r_p =
10  - 50$  $\hkpc$ with  a  line of  sight velocity  difference of  $<
500$~\kms.   Notice  that  pair  counts  based  on  galaxies  selected
according to  $V_{in}$ are  comparable to data  from both the  UZC and
SSRS2  redshift surveys.   In addition,  note the  radically different
slope for the $V_{now}$ line as well as its under-prediction of $N_c$.
The points  along the $V_{in}$  line represent the average  over three
rotations  of the  four realizations  at cuts  ranging from  $V_{in} =
110-350$ in  steps of  20~\kms.  For $V_{now}$  the points  range from
$\Vnow = 110-230$  in steps of 20~\kms. The  observational data points
clearly favor the $\Vin$ model over the $\Vnow$ model.
}
\label{fig:data_compare}
\end{figure}
%

In Figure~\ref{fig:data_compare} we  show the close companion fraction
$\Nc$, derived from galaxies taken from the UZC (solid squares for the
North field and  solid circles for the South) and  the SSRS2 and CNOC2
(solid  triangles) redshift  surveys as  a function  of  galaxy number
density  using  the measured  luminosity  functions  for the  UZC-CfA2
North,   UZC-CfA2  South,   and  SSRS2   surveys  \citep{Marzke_eta94,
Falco_eta99,  Fasano84,  P2000}.   The  CfA2  North  field  originally
covered  a range  of declination  from $8.5^{\circ}  \leq  \delta \leq
44.5^{\circ}$.   For it's Northern  field CfA2  has a  right ascension
range  of $8^h  \leq  \alpha \leq  17^h$, \citet{G&H89,  Huchraetal90,
Huchraetal95}.   The  original CfA2  South  field  covers  a range  of
declination from $-2.5^{\circ} \leq \delta \leq 48^{\circ}$.  The CfA2
survey's  southern field  has a  right ascension  range of  $20^h \leq
\alpha  \leq  4^h$,  \citet{G&H85,  Giovanelli_etal86,  Haynes_etal88,
G&H89b,  Wegner_etal93,  G&H93}.   The  CfA2 North  field  has  $6500$
galaxies  while  the  South   field  has  $4283$  galaxies  originally
cataloged.  These  are volume-limited  samples from $2300  \leq cz\leq
{\rm  H_0} 10^{G({\rm M_{B,lim}})}  $~km~s$^{-1}$.  Here  the exponent
$G({\rm M_{B,lim}}) = (15.5-{\rm  M_{B,lim}}-25)/5$ is a function of a
variable  limiting  magnitude  and  $M_{B}  = 15.5$  is  the  limiting
apparent  magnitude of the  surveys.  We  derive number  densities for
each point by integrating the luminosity function from ${\rm M_{B,lim}
=  -18}$,$-19$, and  $-20$,  respectively to  a  minimum magnitude  of
$M_{B}=-30$.  Note that within the same survey the data points are not
independent. Differences among the  surveys arise, in part, because of
differences in large scale structure, see \S 3.3.

The   results  of   our  model   pair   counts  are   also  shown   in
Figure~\ref{fig:data_compare}  for  both  selection  on  $\Vin$  (open
squares) and selection on  $\Vnow$ (open triangles).  The data clearly
favor the higher pair counts that follow from selecting galaxies based
on the initial maximum circular velocities of their subhalos when they
entered their hosts, $\Vin$.  The $\Vnow$ selection under-predicts the
data by a factor of $\sim 3-5$, depending on the number density.  {\em
We emphasize  that both initial  circular velocity and  final circular
velocity selections  are made from  the same underlying  population of
host dark matter halos, which thus have identical merger histories and
instantaneous merger  rates.}  The differences in  pair counts reflect
only differences  in the evolution of satellite  galaxies within their
host dark matter halos.

%
%
%
\begin{figure}[t!]
\epsscale{0.95}
\plotone{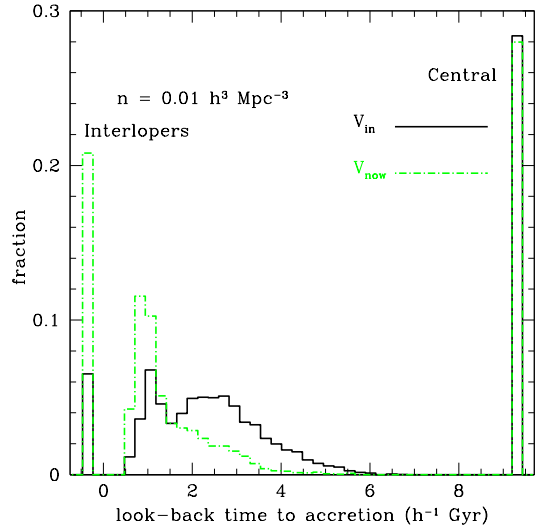}
\caption{
Fraction  of unique  pair-identified  galaxies with  a given  lookback
accretion time. We include each  galaxy identified as part of a unique
$z=0$ pair and  use $\ng = 0.01$ h$^{3}$  Mpc$^{-3}$ with $Vin$ (solid)
and $\Vnow$  (dashed). If  a pair of  galaxies sits in  different host
halos, then each galaxy is assigned an (unphysical) negative accretion
time  and placed  in the  far  left bin  (interlopers). If  a pair  of
galaxies sits in the same host  halo, then each galaxy is assigned the
time it  was accreted into  the host. If  one of the  same-halo paired
galaxies happens to be a central object, we set the lookback accretion
time to be the age of the universe (far right bin).
}
\label{fig:lookback}
\end{figure}
%
%

While the host dark matter halo merger rates are identical between the
$\Vin$   and   $\Vnow$   models,    the   accretion   times   of   the
galaxy-identified  subhalos   in  the  two   cases  are  significantly
different.   Figure  \ref{fig:lookback}   shows  the  distribution  of
lookback accretion times for each object identified in a unique galaxy
pair at $z=0$ for $n_g =  0.01 h^{-3} \Mpc^3$ using $\Vin$ (solid) and
$\Vnow$ (dashed). In this figure  every unique galaxy pair is assigned
two accretion  times, one  for each galaxy  involved in the  pair. For
example, if  there are $N$  close companions then we  include $N(N-1)$
accretion times for those galaxies. If a pair of galaxies is contained
in the same host halo, we assign  each galaxy a time based on its time
of accretion into the host halo.  If one of the same-halo objects is a
central  galaxy, we  let  its ``accretion  time''  be the  age of  the
universe (far right  bin). If the two galaxies  identified as a unique
pair are part  of different halos (interlopers) we  assign each galaxy
an unphysical, negative accretion  time (the far left bin).~\footnote{
Note  that if  a galaxy  is part  of more  than one  unique  pair, its
accretion time  can be  included in this  diagram more than  once. For
example, consider an object with two close companions, one of which is
an  interloper in a  different host  halo and  the other  is contained
within the  same host.  We record a  negative accretion time  for this
galaxy  when we  are  counting the  interloper  pair and  we record  a
non-negative accretion time when counting  it as part of the same-halo
pair.}

The far right bin in Figure~\ref{fig:lookback} shows that $\sim 28 \%$
of  galaxies  in  close  companions  correspond  to  central  objects.
However,  note that  if  we  were enumerating  pairs  rather then  the
galaxies in the  pairs, most pairs on these scales,  in our case $\sim
56\%$, would consist of a satellite and a central object, as one might
expect  from  the  familiar  ``one  halo term''  in  the  halo  model.
~\footnote{Moreover, if  we identify  3-d pairs within  50 kpc  at the
same number density (rather than  in projection) the fraction of pairs
involving a central galaxy grows  to $\sim 76\%$.}  Similarly, $\sim 5
\%$  ($\sim 20$)  of paired  galaxies are  ``interlopers''  sitting in
different  host  halos  (far   left  bin)  for  the  $\Vin$  ($\Vnow$)
choice. As seen in  the central histogram, most unique pair-identified
galaxies are  subhalos. The $\Vnow$-chosen satellite pairs  tend to be
accreted    fairly    recently,   within    the    last   $\sim    0.5
-2$~$h^{-1}$~Gyr. On  the other hand, the $\Vin$  galaxies can survive
longer  within  the  host   potential  without  dropping  out  of  the
observable sample,  and have a  fairly broad range of  accretion times
($\sim  1-4  $~$ h^{-1}$~  Gyr).   More  generally,  we see  that  the
inferred accretion  time distribution  for observed galaxy  pairs will
depend sensitively on galaxy formation assumptions.

\subsection{Cosmic Variance}

Cosmic  variance  is a  major  concern  for  the result  presented  in
Figure~\ref{fig:data_compare}.  The UZC and SSRS2 surveys both contain
structures that are comparable in size  to the surveys as a whole, and
are therefore not ``fair'' samples of the large-scale structure of the
universe.  For  example, the  data clearly reflect  the fact  that the
SSRS2  survey is  underdense relative  to UZC.   Here,  we investigate
whether  the high  observed  pair fractions  could  simply reflect  an
overdensity on a scale comparable to that of the surveys.

\begin{figure}[t]
\epsscale{1.0}
\plotone{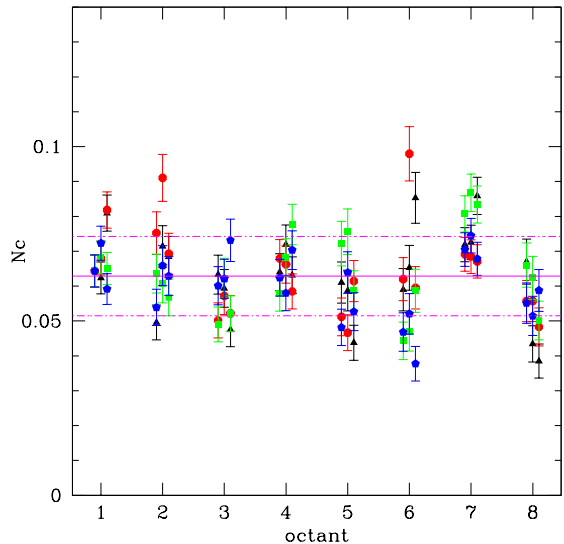}
\caption{   Testing  the   importance  of   cosmic  variance   in  the
observational data sets.   We show $\Nc$ measured in  each octant (1-8
on the horizontal axis) of the  simulation volume for each of the four
Monte-Carlo  subhalo realizations  (four symbol  types at  each octant
number) and  for three  rotations of each  simulation box,  the points
offset from  each octant.  Error bars on  subhalo realizations reflect
Poisson errors on the number  of pairs.  Here we select model galaxies
using a  fixed $\Vin \simeq  162$~\kms, which corresponds to  a number
density {\em  within the full box}  of $n_g =  0.01 h^{3}$ Mpc$^{-3}$.
The mean, over all of the  companions fractions, is given by the solid
line while  the dot-dashed  lines show the  RMS variation,  $0.063 \pm
0.011$.  This mean  value  does not  significantly  differ from  those
produced using only one of the possible lines of sight. }
\label{fig:octvar}
\end{figure}

To examine  the uncertainty in pair  counts due to  cosmic variance in
more  detail we  performed mock  observations of  subsets of  our full
simulation volume  to determine  the variation in  the pair  counts in
volumes    comparable    in    size    to    the    survey    volumes.
Figure~\ref{fig:octvar} shows the variation  in $\Nc$ over eight cubic
regions of the  simulation for each realization, each  $60 \hMpc$ on a
side  for  a  volume  of  $2.16  \times  10^5  h^{-3}$~Mpc$^3$.   This
sub-volume  is within  12\% of  the the  volume-limited UZC  survey to
M$_{\rm B,lim}=-19$.  It is roughly half of the volume for the M$_{\rm
B,lim}=-20$  points for UZC-CfA2  North and  UZC-CfA2 South;  and more
than twice  the volume for  the M$_{\rm B,lim}=-19$ SSRS2  point.  For
each  octant,  we plot  twelve  values of  $\Nc$:  each  of the  three
projections  through  the  box  along with  four  statistical  subhalo
realizations for each projection.  The four symbol types correspond to
each  of  the  four  realizations  and we  have  introduced  a  slight
horizontal shift to reflect  the three sets of rotational projections.
We select model  galaxies by fixing the number density  to $n_g = 0.01
h^{3}$~Mpc$^{-3}$ {\em  within the full  box}.  This corresponds  to a
limit of  $\Vin \gtrsim  162$~\kms.  The reason  for this  approach is
that it mimics what is done  in the observational sample, and uses the
same effective  luminosity function for  the whole volume of  the box.
If we instead choose the same number density within each sub-volume of
the box, the  overall octant-to-octant scatter is reduced  by a factor
of $\sim 2$.  The solid line represents the mean over all $96$ points,
while the  dot-dashed line shows the  RMS variation: $\Nc  = 0.063 \pm
0.011$.  Using  only one projection yields a  nearly identical result.
Error bars on each point are Poisson errors on the number of pairs and
reflect the expected variation in pair counts in the absence of cosmic
variance.  Notice that the scatter  is somewhat larger than what would
be expected from shot noise.

As   we   see   from   Figure   \ref{fig:octvar},   at   maximum   the
octant-to-octant  (large-scale structure)  variation  is $\sim  50\%$.
This  is much  smaller than  the  $\sim 300\%$  variation required  to
reconcile  the $\Vnow$ model  with the  data.  We  also note  that the
scatter within  the local surveys  is much smaller than  the variation
required to  reconcile the $\Vnow$  model with the  observational data
sets.   Nonetheless, repeating  this experiment  with a  larger survey
with better completeness will be required to verify this agreement.

%
%
%
%

\subsection{General Predictions}
\label{sub:pred}

\begin{figure}[t!]
\epsscale{1.0}
\plotone{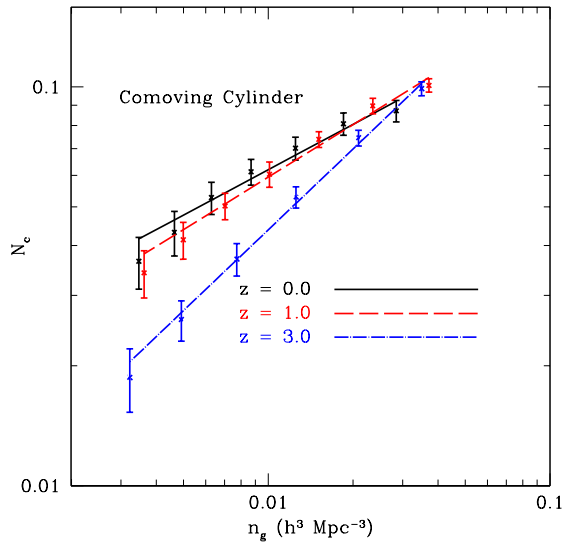}
\caption
{
The  average  companion count  as  a  function of  ($\Vin$-identified)
comoving  volume,  $n_g$,  using  a cylinder with fixed {\em
comoving} volume for companion  identification.  We see that once the
expansion of  the universe  is accounted for  the companion  count per
galaxy is predicted to {\em  decrease} with redshift at fixed comoving
number density (compare to  Figure \ref{fig:Nc_n}, which uses a fixed
{\em physical} cylinder volume in defining $\Nc$).
}  
\label{fig:comov}
\end{figure}

Figure~\ref{fig:Nc_n}  illustrated our  expectation  that  the average
companion number $\Nc$, within a fixed  {\em physical} cylinder volume
increases  with redshift  at   fixed  mean co-moving number   density.
However, this increase does not  reflect the increased merger activity
associated  with hierarchical   structure  formation.  In   fact,  the
predicted evolution is driven primarily by the  choice of a fixed {\em
physical} cylinder to define pairs rather than a selection volume that
is fixed in {\em  comoving}  units.  Figure~\ref{fig:comov} shows  the
identical  statistic computed from  our model galaxy populations using
the selection on $\Vin$ at $z=0, 1$, and  $3$, but this time using the
{\em  comoving}  cylinder  (with  volume   $\propto  (1+z)^3$).    The
virialized radius  of host  halos  {\em at fixed  mass} decreases with
redshift  roughly  as $R_{\rm vir}  \propto   (1+z)$. With  these  new
coordinates we continue to  probe the same  ``fraction'' of each  halo
(of fixed mass) over the entire redshift range.  First, notice that in
comoving units the variation in pair counts with redshift is much more
mild   than   it   is    in   physical   units.    The     counts   in
Figure~\ref{fig:comov} span at most  a factor of  two while the counts
in Figure~\ref{fig:Nc_n} span more than  an order of magnitude over an
identical redshift range.  Second, after normalizing out the effect of
the  cosmological expansion, the  pair  count per galaxy actually {\em
decreases}  with    redshift rather  than   increasing    as shown  in
Figure~\ref{fig:Nc_n}.

\begin{figure*}[t]
\epsscale{1.1}
\plotone{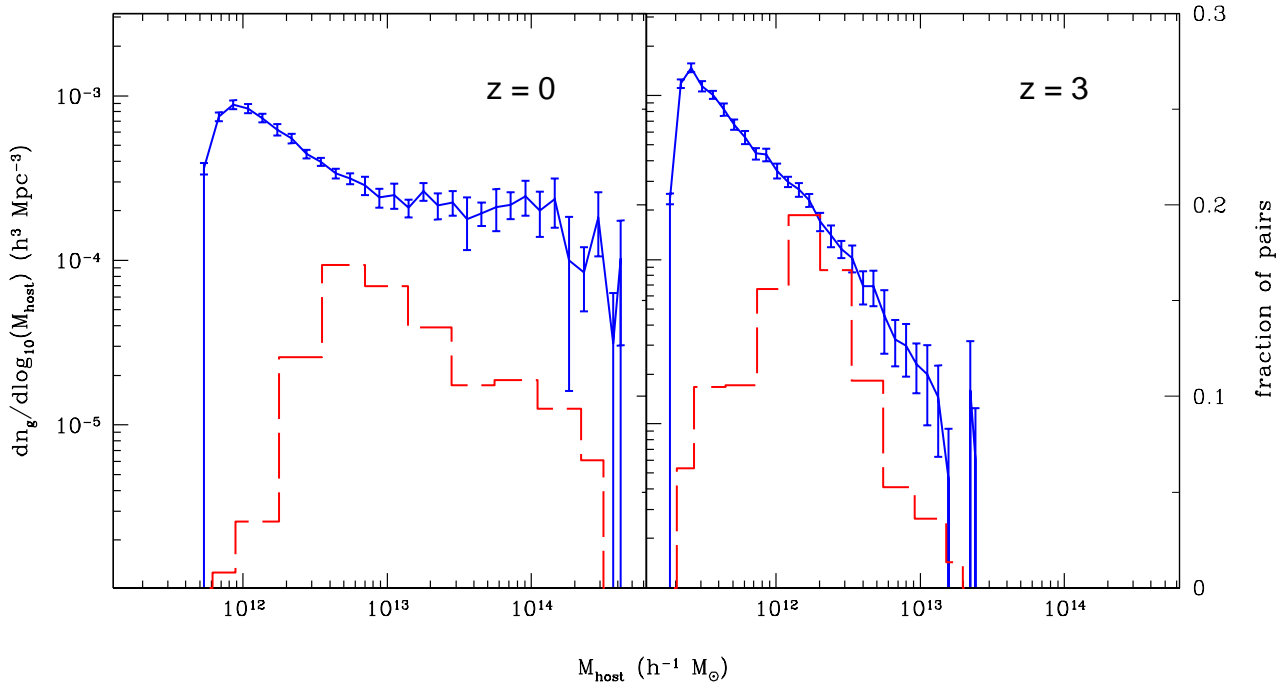}
\caption
{
The total galaxy  number density distribution  as a function of  galaxy host
halo mass  (solid lines,  left axis) compared  to the  distribution of
host  halo masses  for identified  galaxy pairs  (dashed  lines, right
axis).   In each case we use the $\Vin$ model with
$n_g =  0.01$ h$^3$ Mpc$^{-3}$.
Paired galaxies are biased to sit in larger halos than their ``field''
counterparts.  At high redshifts, the larger host halos
become increasingly rare, and this is reflected in distribution of host
masses for paired galaxies.
}
\label{fig:mass_hist}
\end{figure*}

The primary reason why the companion fraction does not evolve strongly
with redshift is that the  number density of host halos massive enough
to host more than one  bright galaxy decreases at high redshift.  This
is shown  explicitly in Figure~\ref{fig:mass_hist}, where  we plot the
distribution of host halo masses  for galaxies (solid lines with error
bars, left axis) at $z=0$ and $3$  using $\Vin$ to fix $n_g = 0.01 h^3
{\rm Mpc}^{-3}$.  The dashed line  in each panel (right vertical axis)
shows the distribution of host  halo masses for {\em pairs} identified
in the same sample of galaxies.  At all redshifts, pairs are biased to
sit in  the most massive host  halos.  At high  redshift massive halos
become rare, and the pair fraction is reduced accordingly.

%
%
%

%
%
%

%
%
\begin{figure}[t]
\epsscale{1.05}
\plotone{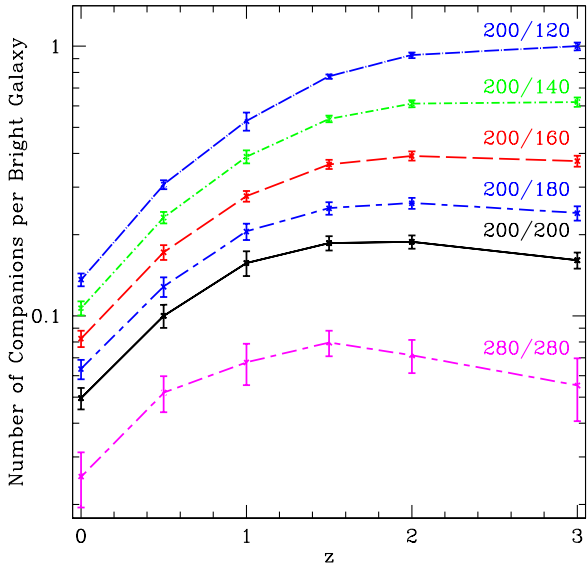}
\caption
{  The average  number  of close  companions  per bright  galaxy as  a
function of  redshift defined using  our standard definition  of close
pair with  a fixed {\em physical}  separation.  In all  but the lowest
line, ``Bright'' objects  are defined as objects with  a $V_{in}> 200$
km s$^{-1}$.   The pair fraction for ``Bright-Bright''  pairs is shown
by   the   solid  line   (marked   ``200/200'')   and  even   brighter
``Bright-Bright''  pairs  (using $\Vin  >  280$  \kms)  as the  lowest
long-short-dashed line (marked ``280/280'').  The other lines show the
average ``faint'' companions per bright galaxy.  We identify ``Faint''
objects  using cut-offs  of  $V_{in} \geq  X$  km s$^{-1}$  with $X  =
180,160,140,$  and 120.   Not  only is  the  faint companion  fraction
systematically  higher than  the  bright companion  fraction, it  also
evolves more  strongly with redshift --  a factor of $\sim  10$ out to
$z\sim 2$ for the $200/120$ case  compared to a factor of $\sim 3$ for
the  $200/200$  case.    Note  that  the  brightest  ``Bright-Bright''
companion count shows a definite drop beyond $z \sim 1.5$.  }
\label{fig:BBFB}
\end{figure}
%
%

Given that halos  with multiple {\em bright} galaxies  are expected to
become increasingly rare at  high redshift, faint companion counts may
provide  a   more  useful  probe.    Figure~\ref{fig:BBFB}  shows  the
evolution in the average number of ``Faint'' companions per ``Bright''
galaxy  as  a function  of  redshift  using  our standard  {\em  fixed
physical}  separation  criteria for  defining  close  pairs.  The  top
(dot-dashed)  line  shows  the   average  number  of  ``Faint''  close
companions  with $\Vin  >  120$~\kms (${\rm  M_B}  \lesssim -17.3$  at
$z=0$) around  ``Bright'' galaxies with $\Vin >  200$~\kms (${\rm M_B}
\lesssim -18.9$ at $z=0$) as a  function of $z$.  The four lines below
this  restrict companions  to be  ``brighter'' galaxies  with  $\Vin =
140$, $160$,  $180$, and finally $200$  \kms.  The last  of these four
(solid  line)  represents  our  standard  ``Bright-Bright''  companion
count, $\Nc$.  Finally, the lowest  line shows the evolution in ``Very
Bright-Very Bright'' companion counts  using $\Vin = 280$~\kms (marked
``280/280'').    As  expected,   the  faint   companion   fraction  is
systematically higher than the  bright companion fraction, but it also
evolves  more  strongly  with  redshift.   On  average,  almost  every
``Bright galaxy'' (with  $\Vin > 200$~\kms) at $z >  2$ is expected to
host a ``faint'' companion  (with $\Vin > 120$~\kms).  The ``200/120''
Bright-Faint companion  count rises  by almost a  factor of  $\sim 10$
between $z=0$ and $z=2$, while the  increase is only a factor of $\sim
3$  for the  $200/200$ case.   Indeed the  brightest ``Bright-Bright''
companion  count  (280/280) actually  declines  beyond  $z \sim  1.5$.
These  trends  reflect both  the  decline  of  massive halos  at  high
redshift  and  the competition  between  accretion  and disruption  of
subhalos, which favors accretion at high redshift.

\section{Interpretations and the Halo Model}
\label{sec:interp}

As demonstrated above, predicted (physical) close pair counts do not evolve
rapidly with redshift, even at fixed global number density.  
This result is driven by a competition between 
increasing merger rates and decreasing
massive halo counts:
while the number of galaxy pairs per host
halo increases with $z$ (as the merger rate increases), the number of
halos massive enough to host a pair decreases with $z$.
In this section we will
work towards a more precise explanation and compare our two models for
the connection between galaxy light and the subhalo properties $\Vin$
and $\Vnow$, by investigating their Halo Occupation Distributions
(HOD).

The  HOD is the  probability, $P(N|M)$, that $N$  galaxies meeting some
specified selection criteria reside within the virial radius of a host
dark matter  halo of mass  $M$.  In what  follows, we provide  a brief
introduction to the  HOD as used in galaxy  clustering predictions and
use a  ``toy'' HOD  model to show  that galaxy close  companion counts
provide an  important constraint  on the HOD  and the  distribution of
galaxies within halos \citep[see also][]{bws:02}.  We then
quantify our two subhalo models, $\Vin$ vs. $\Vnow$, in terms of these
inputs and  present this as a  general constraint on the  kind of HODs
required to match the current pair count data.

%
%
%
\begin{figure*}[t]
\epsscale{1.1}
\plottwo{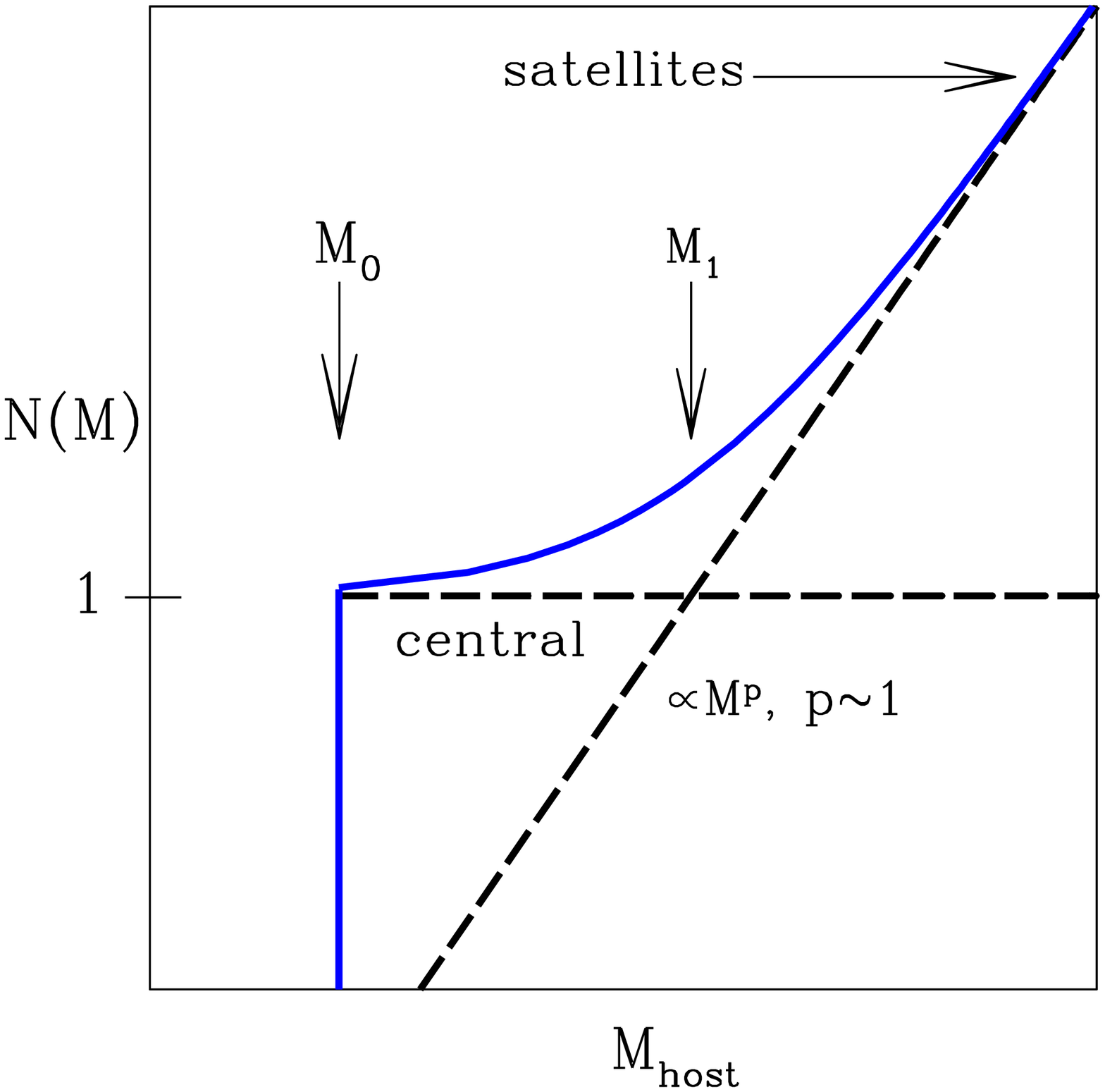}{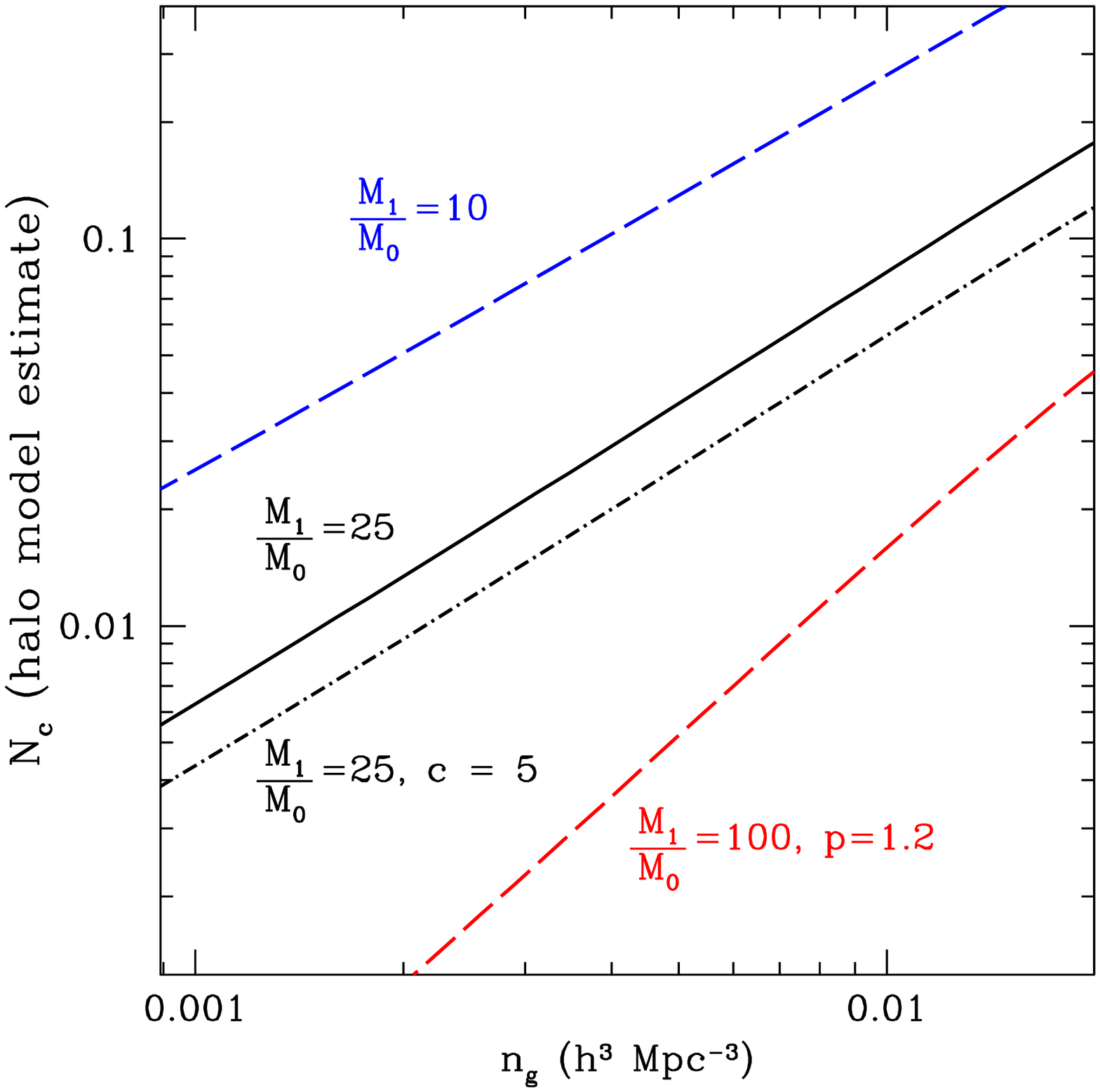}
\caption
{ {\em Left:} A cartoon  representation of an input halo model $N(M)$,
where the  plateau at  $N(M) = 1$  represents central objects  and the
power-law piece,  $\propto M^p$, represents  satellite galaxies around
the central galaxy.   The ratio of $M_0$ to  $M_1$ sets the importance
of satellite galaxies compared to ``field'' galaxies -- the larger the
ratio, the more  $n_g$ will depend on central,  $N=1$, galaxies.  {\em
Right:}  The  pair  fraction  as  estimated by  a  simple  halo  model
calculation (see Equation  \ref{eqn:pairs} and associated discussion).
Each  line  is  labeled  with  its assumed  ratio  of  $M_0/M_1$.   As
expected, models with a  larger $M_0/M_1$ ratios have fewer companions
per galaxy because field galaxies  are emphasized.  All lines assume a
power-law  index of  $p=1.0$ except  for the  lowest dashed  line with
$p=1.2$.  In  this case,  the relation with  $n_g$ is  steeper because
satellite galaxies tend to be more plentiful in lower-mass host halos.
The  line labeled  ``$c=5$''  has a  more  diffusely packed  satellite
distribution and as a result produces fewer close pairs with all other
parameters held fixed.}
\label{fig:hodgen}
\end{figure*}
%
%
%

\subsection{Pair counts via the halo model}
\label{sub:halomodel}

The  halo  model  is   a  framework  for  calculating  the  clustering
statistics  of galaxies  by assuming  that  all galaxies  lie in  dark
matter    halos   \citep[e.g.][]{seljak:00,peacock_smith:00,
scoccimarro_etal:01}.   This  approach relies  on  the  fact that  the
clustering properties  and number densities of host  dark matter halos
can be predicted accurately, and  uses a parameterized HOD to make the
connection between dark matter halos and galaxies.

Consider  an HOD  which consists  of a  ``central''  and ``satellite''
population  of  galaxies  orbiting   in  the  same  dark  matter  halo
\citep[e.g.][]{kwg93,Kravtsov04a,tinker:05,cm:05}.   The total  HOD is
then described by two distributions $P(N_{c}|M)$ and $P(N_{s}|M)$ with
individual mean values $\Ncen = \int N_c P(N_c|M) {\rm d}N_c$ and $\Ns
= \int N_s  P(N_s|M) {\rm d}N_s$ respectively.  Let  us assume that we
can write the first moment of $P(N|M)$ (the average number of galaxies
per host halo of mass $M$) as
\begin{eqnarray}
\N  = \Ncen + \Ns = \left\{\begin{array}{cll}
	&  1 + \left( \frac{M}{M_1} \right)^p  & M \ge M_0 \\
        &  0  &  M < M_0,\\	   \end{array} \right.
\label{eqn:toyhod}
\end{eqnarray}
where  $M_0$  is  the  minimum  host  mass large  enough  to  host  an
observable  galaxy, $M_1$  is the  typical host  mass which  hosts one
observable satellite  galaxy, and $p$  is a power which  describes the
scaling of satellite galaxy  number with increasing host mass.  Though
simplified, this general  form with $p\approx 1$ is  motivated by both
analytic    expectations    \citep[e.g][Z05]{Wechsler01}    and    the
expectations of simulations \citep[e.g.][]{Kravtsov04a,tinker:05}, and
produces large-scale  clustering results  that are in  broad agreement
with  data  selected  over  a  range  of  luminosities  and  redshifts
\citep{conroy_etal05}. These  results also  motivate us to  assume that
the  satellite HOD is  given by  a Poisson  distribution and  that the
central  piece  is represented  by  a  sharp  step function  (in  this
approximation)   at  $M   \ge  M_0$.    The  left   panel   of  Figure
\ref{fig:hodgen} shows a cartoon representation of this HOD.

With $P(N|M)$ given, the number  density of galaxies can be written as
an integral of $\N$ over halo mass,
\begin{equation} n_g =
\int_{M_{0}}^{\infty}dM
\frac{dn}{dM} \N,
\label{eqn:ng}
\end{equation}
where, $dn/dM$ is  the host dark matter halo  mass function.  As shown
in Figure~\ref{fig:massfunc}, the host halo mass function is a rapidly
declining  function of  mass.  This  implies that  the  overall galaxy
number density  is dominated  by ``field'' galaxies  in host  halos of
mass $M_1 \lsim M \lsim M_0$, where $\N \simeq 1$.

%
%
%
\begin{figure*}[t]
\epsscale{1.05}
\plotone{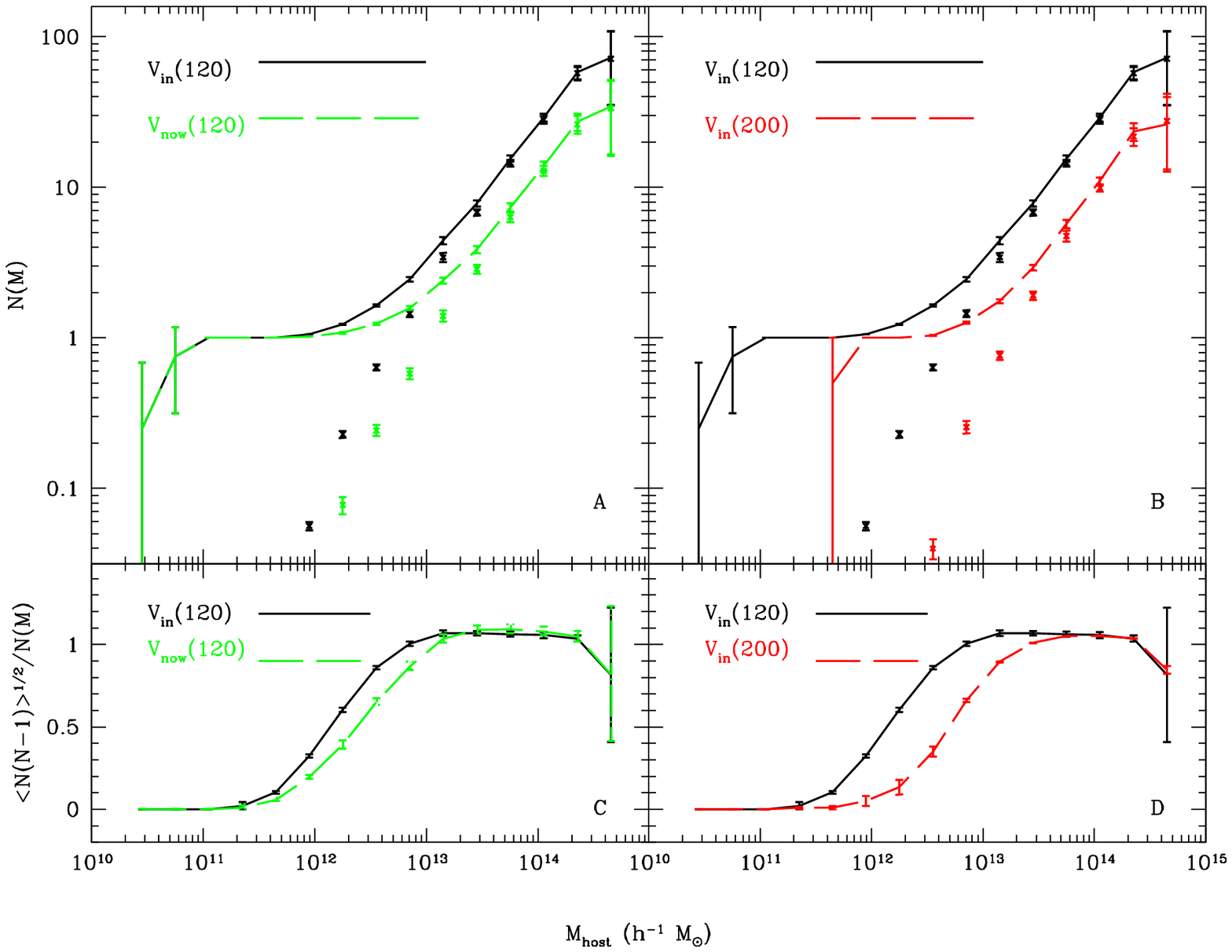}
\caption
{ Top: The  average number of galaxies per host halo  as a function of
host  halo   mass.   Bottom:  Expression   for  the  pairs   per  halo
$<N(N-1)>^{1/2}/N(M)$  (relative  to  the  Poisson expectation)  as  a
function of  host halo mass.  In  the upper panels,  points with error
bars reflect  satellite galaxies and  lines show the  total (satellite
plus central)  galaxy count  per halo.  The  left figure  compares the
$\Vin$  model  (solid)  and  $\Vnow$  model (dashed)  using  the  same
velocity cut  to define the  galaxy population (120 \kms).   The right
panel compares  the $\Vin$ model  at two different velocity  cuts (120
and 200 \kms).  The pair count per halo is well described by a Poisson
distribution for  the satellites and  substantially sub-Poissonian for
the central objects.  Pair counts thus become important only for $N(M)
\gsim 2$, and will be dominated by massive halos.  }
\label{fig:HOD1}
\end{figure*}
%
%
%

On the other hand, for  small separations, the number density of close
pairs, $n_p$, will be dominated by pairs contained within single halos
(Figure~\ref{fig:Purity}  illustrates  that,  for  the  $\Vin$  model,
$\lsim 10\%$ of close pairs reside in separate host halos according to
typical close-pair definitions  out to $z \leq 3$).   Ignoring for the
moment the precise velocity selection criteria, we can approximate the
number density  of pairs  with physical projected  separations meeting
some range $r_1 < r < r_2$ as
\begin{equation}
n_{\rm p} \simeq n_{\mathrm{p,1h}} =
\frac{1}{2} \int_{M_{\rm min}}^{\infty}dM
\frac{dn}{dM} \NN_M F\left(r_{1,2}|M\right),
\label{eqn:pairs}
\end{equation}
where $\NN_M$  is the second moment  of the HOD  and $F(r_{1,2}|M)$ is
the fraction  of galaxy pairs that have  projected separations between
$r_1$ and $r_2$  within a dark matter halo of  mass $M$.  The function
$F$ provides an $\sim$ order unity covering factor determined by the
distribution of galaxies within halos~\footnote{More precisely, $F$ is
determined  by  convolving   the  projected  density  distribution  of
galaxies  within halos  with  itself.}.  Notice  that  since $\NN_M  =
\Ns^2$ and $\Ns \rightarrow 0$ for  $M < M_1$, the pair number density
$n_p$  (unlike  $n_g$)  explicitly  selects against  the  mass  regime
dominated by ``central galaxies'' and favors halo with $M > M_1$.

%
%
\begin{figure}[t!]
\epsscale{1.0}
\plotone{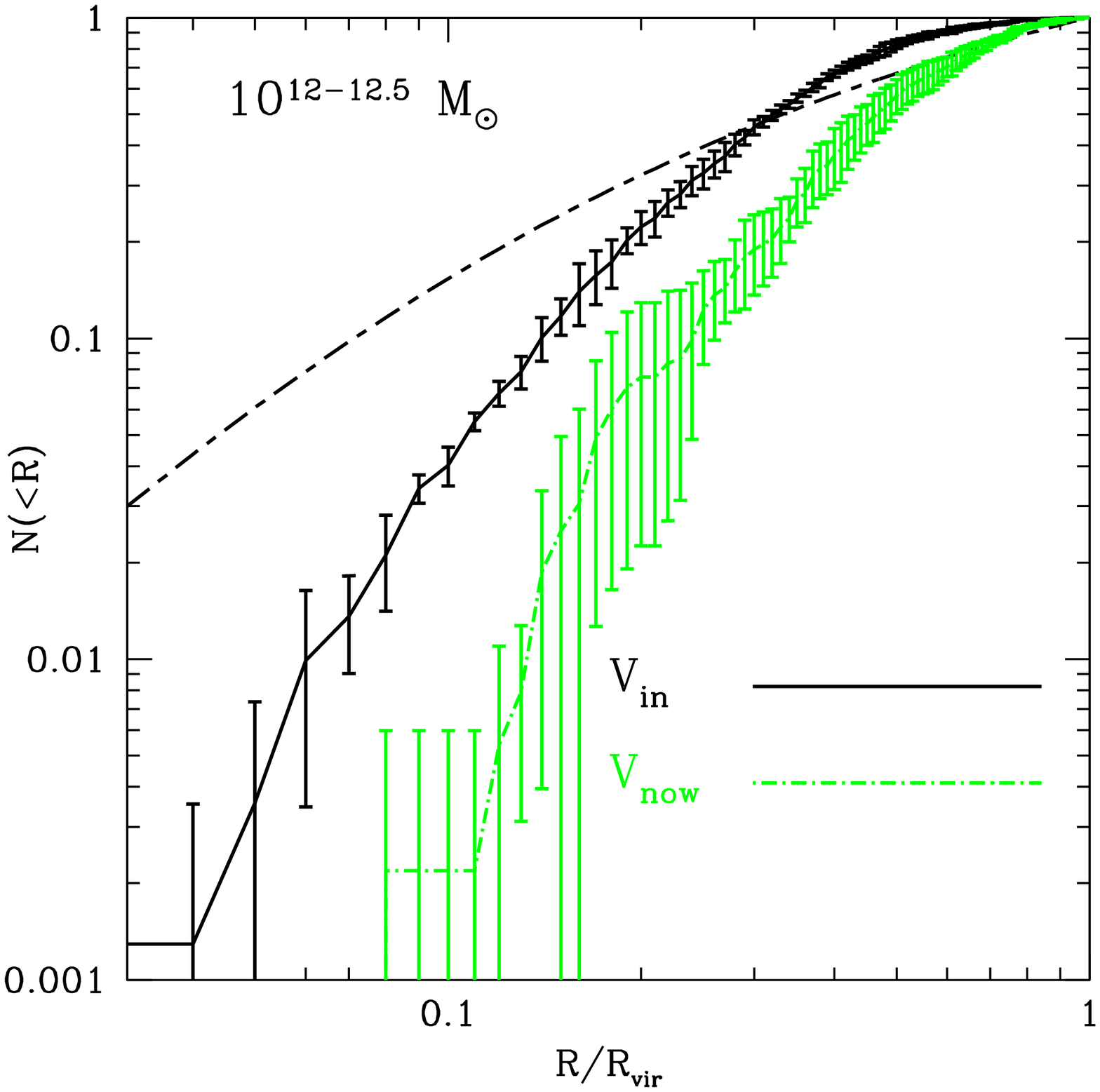}
\caption
{ The normalized cumulative number of satellite galaxies as a function
 of  radius  from  the  host  halo  center  for  host  halos  of  mass
 $10^{12-12.25}  \hMsun$ ($\Rvir  \simeq 204  \hkpc$).  The  solid and
 dash-dot  lines  correspond  to   $\Vin$  and  $\Vnow  =  120$  \kms,
 respectively.  The cumulative mass profile for a typical halo of this
 mass is  shown by the dot-dash  line.  Error bars  are RMS variations
 over the four Monte-Carlo subhalo realizations.  }
\label{fig:NC}
\end{figure}
%
%
%

This simple and intuitive  formalism immediately provides insight into
the nature  of the companion  fraction, $N_c \equiv 2  n_{\rm p}/n_g$.
In  the  right-hand  panel  of  Figure \ref{fig:hodgen}  we  show  the
predicted  value   of  $\Nc$  vs.   $n_g$  as  calculated   using  the
(approximate) equations  \ref{eqn:ng} and \ref{eqn:pairs}  with $r_1 =
10 \hkpc$ and  $r_2 = 50 \hkpc$ for several  choices of HOD parameters
defined  in Equation \ref{eqn:toyhod}.   In all  cases we  assume that
galaxies are distributed in  dark matter halos following NFW profiles.
The upper  long-dashed line and solid  line assume $M_1/M_0  = 10$ and
$25$ respectively, use $p=1$ in  the satellite $\Ns$ piece, and assume
the    NFW    concentration-mass    relation,   $c(M)$,    given    in
\cite{Bullock01a}. The dot-dashed  line is the same as  the solid line
except we fix the NFW  concentration to a relatively low value, $c=5$,
for all halos.  Finally, the lower dashed line assumes $M_1/M_0 = 100$
with $p=1.2$ and the standard concentration relation.

From  Figure  \ref{fig:hodgen}  we  see   that  as  the  size  of  the
``plateau''  region increases (characterized  by increasing  the ratio
$M_1/M_0$)  the galaxy  number  density is  increasingly dominated  by
central galaxies  and the pair fraction  goes down.  Also,  even for a
fixed  HOD,  $\Nc$ decreases  as  galaxies  are  packed together  more
diffusely in  halos (characterized by lowering $c$).   Finally, as the
satellite  contribution  steepens  (higher  $p$), $\Nc$  increases  at
higher number densities because lower mass host halos begin to contain
multiple galaxies.

As illustrated  by this  toy-model exploration, pair  count statistics
should provide an  important constraint on galaxy HODs  as well as the
distribution  of galaxies  within halos.   The constraint  could prove
particularly powerful  when combined with larger-scale  ($r \gsim 100$
kpc)   two-point  correlation   function  constraints   on   the  HOD.
Correlation function constraints  generally suffer a severe degeneracy
between $P(N|M)$ and $F(r|M)$, but  this degeneracy may be broken when
coupled with {\em close} ($r \lsim 50$~kpc) pair fraction constraints.

\subsection{Discussion of Model Results}

The  halo  model serves  as  a  useful  framework for  discussing  and
explaining our simulation  model results.  Figure \ref{fig:HOD1} shows
the first and second moments  of our simulation model HODs.  The upper
panels show $N(M)$ and the lower panel shows the number of pairs per
halo compared  to the Poisson  expectation, $<N(N-1)>^{1/2}/<N>$.  The
left  side of the  figure shows  how the  HOD changes  as we  vary the
selection criteria, with $\Vin  > 120$\kms subhalos represented by the
solid line  and $\Vnow > 120$\kms  subhalos shown by  the dashed line.
As must  be the case, the  central galaxy term (analogous  to $M_0$ in
our  toy  example above)  is  unaffected  by  this choice,  while  the
satellite component  (points) is  enhanced when galaxies  are selected
using $\Vin$.  This has the effect of extending the ``plateau'' region
in the $N(M)$ function and  is qualitatively similar to increasing the
$M_1/M_0$ ratio in  our toy-model discussed above.  In  the right hand
panel we show $N(M)$ for  two different velocity cuts (120 compared to
200 \kms) both using $\Vin$ as the velocity of relevance.  Here we see
that {\em both}  the satellite term and the  central term are affected
in  this case  ($M_0$ becomes  larger as  more massive  satellites are
selected).   Note that  as expected  from the  previous  discussion of
satellite vs.  central galaxy  statistics, the average number of pairs
of galaxies  per halo tracks  the Poisson expectation for  host masses
where  $N(M) \gsim  2$  and becomes  substantially sub-Poissonian  for
$N(M) \lsim  2$.  As discussed in reference  to Equations \ref{eqn:ng}
and \ref{eqn:pairs}, this means that while the total number density of
{\em galaxies} (with, e.g., $\Vin \gsim 120$\kms) will be dominated by
host halos where $N(M) \simeq 1$  or with $M \simeq M_0 \simeq 10^{11}
\hMsun$, the total number density  of {\em close galaxy pairs} will be
dominated by much more massive host halos $M \gsim M_1 \simeq 5 \times
10^{12} \hMsun$ (where $N(M) \gsim 2$).

In addition  to its  effect on $P(M|N)$,  whether the galaxy  light is
more  accurately traced  by $\Vnow$  or  $\Vin$ will  also affect  the
relative {\em  distribution} of satellite galaxies  within host halos.
Recall  that  $\Vnow$  selects  ``observable''  galaxies  based  on  a
subhalo's {\em  current} maximum circular velocity  and $\Vin$ selects
galaxies based on the subhalos' circular velocity when it was accreted
into the system  (these might mimic models where  galaxies are more or
less  affected  by  mass  loss  or surface  brightness  dimming  after
accretion).   As shown  in Figure  \ref{fig:NC}, the  $\Vin$ selection
results  in a  satellite galaxy  distribution that  is  more centrally
concentrated (and  thus has more close  pairs per halo)  than does the
$\Vnow$  selected  sample.   The  figure shows  cumulative  number  of
satellites relative to the total number of satellites as a function of
radius, scaled  to the virial radii  of the hosts, from  the host halo
center using host  halo masses of $\Mvir =  10^{12-12.5} \hMsun$.  The
solid line shows this for the $\Vin$ case and the dashed line reflects
the $\Vnow$ case.  For reference,  the dot-dashed line above shows the
cumulative {\em mass}  distribution for a typical dark  matter halo of
this size  \citep{Bullock01a}.  The $\Vnow$ case shows  the effects of
enhanced  satellite  destruction  near  the center.   This  difference
certainly  affects  predictions  for  pair counts,  however  the  main
difference between  $\Nc$ counts  in our two  models is driven  by the
differences in their HODs.

\begin{figure}[!t]
\epsscale{1.0}
\plotone{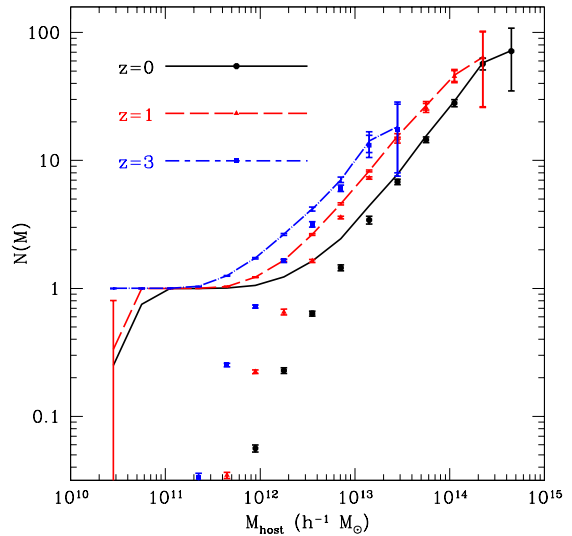}
\caption
{
The average number of galaxies per  host dark matter halo with $\Vin >
120$\kms shown at $z=0$, 1, and 3.  The points reflect satellite halos
while the lines include all galaxies.
}
\label{fig:HOD2}
\end{figure}

Finally  we turn  to  the  evolution of  the  companion fraction  with
redshift.   As discussed in  the previous  section, the  average close
companion  count per  galaxy is  expected to  evolve only  weakly with
redshift  (e.g. Figure \ref{fig:DEEP2_compare})  and even  declines at
fixed comoving number density when pairs are identified within a fixed
comoving separation (Figure \ref{fig:comov}).  This occurs even though
the  merger rate {\em  per host  dark matter}  halo increases  at high
redshift.  This  increase in  merger rate is  indeed reflected  in the
HOD.   As  shown  in  Figure  \ref{fig:HOD2}, the  average  number  of
galaxies per halo {\em at fixed host mass} increases systematically at
high  redshift.  However,  from Equation  \ref{eqn:pairs},  the number
density  of  pairs  depends both  on  the  number  of pairs  per  halo
($<N(N-1)>  \sim  \Ns^2$)  and   the  number  density  of  host  halos
($dn/dM$).  In Figure \ref{fig:HOD2},  for example, $N(M)$ is a factor
of $\sim  2$ higher at  $z\sim 3$ at  the host mass where  pair counts
become  non-negligible $M_{\rm host}  \sim 10^{12}  \hMsun$.  However,
host halos  of this  size are  a factor of  $\sim 5$  rarer at  $z= 3$
compared to $z=0$ (Figure \ref{fig:massfunc}).  This results in a mild
overall decline in the number of same halo pairs at high redshift.

The  HOD formalism  also  helps us  understand Figure  \ref{fig:BBFB},
where  we find  that  the  average number  of  ``Faint'' galaxies  per
``Bright''   galaxy    increases   more   rapidly    than   does   the
``Bright-Bright''  fraction   with  redshift.   The   more  pronounced
evolution comes about  because ``Faint'' satellite selection decreases
the  characteristic   host  mass   required  for  a   satellite  ($M_1
\downarrow$).   At  the same  time,  we  focus  on ``Bright''  central
galaxies, which keeps the minimum host halo mass for hosting a central
galaxy  ($M_0$) roughly fixed.   As $M_1/M_0$  decreases, we  are more
likely to see an increased companion count (Figure \ref{fig:hodgen}).


\section{Conclusions} \label{sec:discussion}

In this paper  we have investigated the use of close  pair counts as a
constraint  on hierarchical  structure  formation using  a model  that
combines   a  large   ($120  \hMpc$   box)  \lcdm   N-body  simulation
\citep{Allgood05}   with  a   rigorous   analytic  substructure   code
\citep{Zentner05} in  order to  achieve robust halo  identification on
small  scales  within a  cosmologically-relevant  volume.  We  measure
close  pairs  in our  simulation  in exactly  the  same  way they  are
observed in the real universe and  use these close pair counts to test
simple, yet  physically well-motivated models  for connecting galaxies
with dark matter  subhalos.  We explore two models  for the connection
between  galaxies and their  associated dark  matter subhalos:  one in
which satellite  galaxy luminosities  trace the {\em  current} maximum
circular velocities of their subhalos,  $\Vnow$, and a second in which
a satellite galaxy's luminosity is tightly correlated with the maximum
circular velocity  the subhalo  {\em had} when  it was  first accreted
into  a host halo,  $\Vin$.  The  latter case  corresponds to  a model
where galaxies  are much  more resilient to  mass loss than  are their
dark matter halos.

A highlighted summary of this work and our conclusions are as follows:

\begin{itemize}

\item  We showed that  \lcdm models  naturally reproduce  the observed
  ``weak'' evolution  in the close  companion fraction, $\Nc$,  out to
  $z\sim 1.1$ as reported by the DEEP2 team \citep{Lin04}.  The result
  is driven  by the fact  that while merger  rates (and the  number of
  galaxy pairs per halo) increase towards high redshift, the number of
  halos massive  enough to  host a bright  galaxy pair  decreases with
  $z$.

\item We used the SSRS2 and  UZC surveys to derive companion counts as
  a function  of the underlying galaxy population's  number density at
  $z=0$  and  showed  that  the  relatively  high  companion  fraction
  observed favors  a model where  galaxies are more resistant  to mass
  loss  than are  dark  subhalos ($\sim$  the  $\Vin$ model).   Cosmic
  variance  is   unlikely  to  be  qualitatively   important  in  this
  conclusion,  although  verifying  the  result  requires  a  complete
  redshift survey with a larger cosmic volume.

\item We  argued that  the close luminous  companion count  per galaxy
  ($\Nc$) {\em  does not}  track the distinct  {\em dark  matter halo}
  merger  rate.  Instead it  tracks the  {\em luminous  galaxy} merger
  rate.  While a  direct connection between the two  is often assumed,
  there is  a mismatch because  multiple galaxies may occupy  the same
  host  dark   matter  halo.   {\em   The  same  arguments   apply  to
  morphological identifications of merger remnants}, which also do not
  directly probe the host dark halo merger rate.

\item We showed that while close pair statistics provide a poor direct
  constraint  on   halo  merger  rates,  they   provide  an  important
  constraint  on the  galaxy Halo  Occupation Distribution  (HOD).  In
  this way, pair  counts may act as a general  tool for testing models
  of galaxy formation on $\lesssim 100$ ~kpc scales.

\end{itemize}

%
%
%
%
%

Our   results  are   not  driven   by  the   effects  of   false  pair
identification.   We showed that  standard techniques  for identifying
close pairs via close projected separations and line-of-sight velocity
differences do quite well in identifying galaxy pairs which occupy the
same  dark matter halos  (Figure \ref{fig:Purity}).   Typically $\gsim
90\%$ of  pairs identified  in this manner  occupy the same  host dark
matter halos out to a redshift of $\sim 2$ using the $\Vin$ model.  It
should be noted  that this fraction does depend on  the details of the
association  between subhalos and  galaxies. The  striking differences
between  our two  toy models  provides evidence  of  this.  Typically,
$\sim 20 \%$ of pairs occupy different halos in the $\Vnow$ model.

In conclusion, current close pair  statistics are unable to rule out a
simple scenario where galaxies are associated in a one-to-one way with
dark  matter subhalos  (in particular  our $\Vin$  association)  in the
current  concordance \lcdm\ model.   Better statistics,  more complete
surveys, and  the ability to divide  samples by color  or stellar mass
indicators will  help refine this  model and explore  galaxy formation
unknowns on $\lesssim 100$ ~kpc scales.  Enumerating {\em faint} close
companion counts  around bright galaxies  (Figure \ref{fig:BBFB}) will
also provide a new and powerful test.

\bigskip

\acknowledgments 

The simulation  was run  on the Seaborg  machine at  Lawrence Berkeley
National  Laboratory (Project  PI:  Joel Primack).   We thank  Anatoly
Klypin for  running the simulation and  making it available  to us. We
thank Jeff  Cooke, Asantha Cooray, Margaret  Geller, Manoj Kaplinghat,
Andrey  Kravtsov,  and Jeremy  Tinker  for  useful conversations.   We
acknowledge   Darrell  Berrier   for  valuable   advice   on  analysis
software. JCB  and JSB  are supported by  NSF grant  AST-0507916; JCB,
JSB, EJB,  and HDG  are supported  by the Center  for Cosmology  at UC
Irvine.  ARZ is funded by the Kavli Institute for Cosmological Physics
at The  University of Chicago  and by the National  Science Foundation
under grant NSF PHY 0114422.   RHW is supported by NASA through Hubble
Fellowship  grant  HST-HF-01168.01-A awarded  by  the Space  Telescope
Science Institute.

\appendix

\section{Pair Fraction as a Function of Number Density}

%
%
%
\begin{figure*}[t!]
\epsscale{1.0}
\plotone{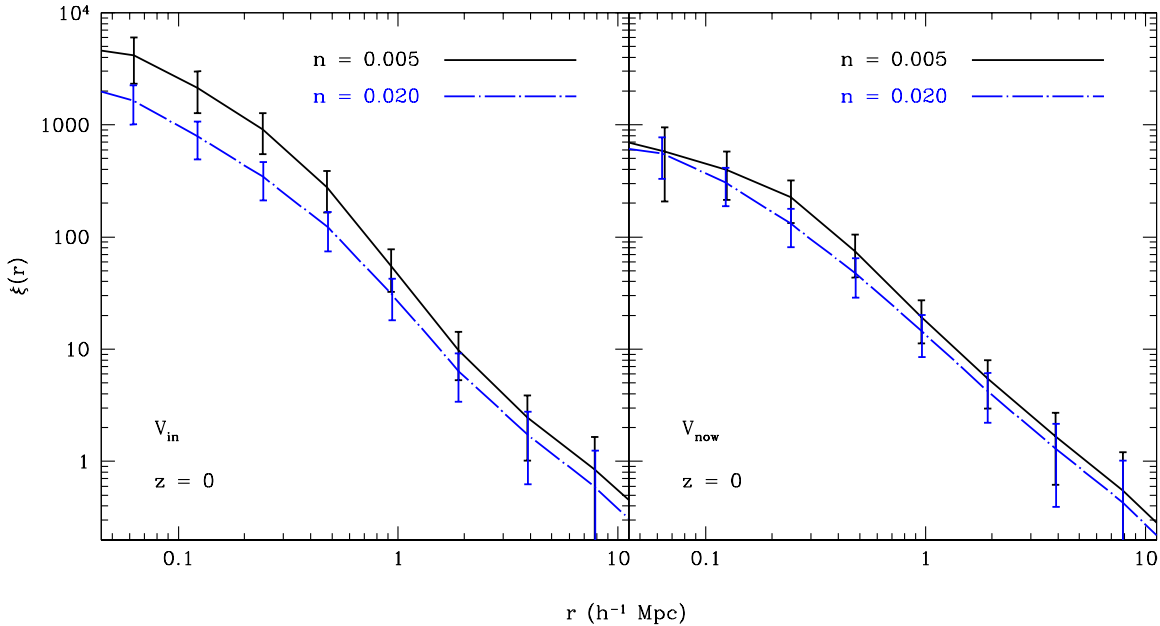}
\caption{ Model  $z=0$ correlation functions at  two number densities.
The left panel  shows number densities defined using  a cut on $\Vnow$
and the right panel  uses $\Vin$.  Error-bars represent the quadrature
sum of the  standard deviation of the four  model realizations and the
cosmic variance computed by jackknife re-sampling of the eight octants
of the volume.  }
\label{fig:V_2pt}
\end{figure*}
%
%

One of  the important  findings in our  analysis is that  the observed
companion  fraction of  galaxies should  increase with  the underlying
galaxy  number  density (see,  e.g.  Figure  \ref{fig:Nc_n}).  We  can
understand  the tendency  for $\Nc$  to increase  with $n_g$  at fixed
redshift  by considering  the  number  of pairs  expected  in a  fixed
(physical) cylinder  of volume $V$  (e.g., with length $\ell  \simeq 2
\Delta v H_0^{-1} \simeq 10 h^{-1}$~Mpc and radius $r \simeq 50$~kpc).
Let us start with the  idealization of an uncorrelated sample.  For an
uncorrelated galaxy  population, with mean  number density $n_g$  in a
large  fiducial volume  $V_0$, the  total  number of  galaxies in  the
sample is clearly \beq N_{\rm gal} = n_g V_0.  \eeq The average number
of companions per galaxy is $N_p = n_g V$, so the number of companions
in the entire sample is simply  $n_g^2\, V \, V_0$.  Therefore, for an
unclustered galaxy  population, the  average number of  companions per
galaxy should increase linearly with galaxy number density, \beq \Nc =
n_g\, V  \propto n_g.  \eeq Extending  this argument to the  case of a
clustered galaxy population is straightforward: \beq N_p = n_g V [ 1 +
\bar{\xi}],  \eeq where  $\bar{\xi} \equiv  V^{-1} \int_V  \xi(r) {\rm
d}V$ is the average of the correlation function over the volume of the
cylinder.  The  total number  of companions in  the sample is  then is
$n_g^2 [1 + \bar{\xi}] V V_0$ and the companion fraction is \beq \Nc =
n_g V  [1 + \bar{\xi}] \approx  n_g V \bar{\xi}, \eeq  where we assume
$\xi(r) \gg 1$ at all relevant scales.  There are both theoretical and
direct  observational reasons  to believe  that $\xi$  should increase
with  decreasing  number  density.   If  we take  a  simple  power-law
description $\bar{\xi}  \propto n_g^{-\alpha}$  then we have  \beq \Nc
\propto n_g^{1-\alpha}.   \eeq Therefore, quite generally,  as long as
$\alpha < 1$, the companion  fraction should be an increasing function
of galaxy number density.

The $z=0$  correlation function for  our two simulation models  at two
characteristic galaxy number densities  is shown in the left ($\Vnow$)
and  right  ($\Vin$)  panels  of Figure  \ref{fig:V_2pt}.   Note  that
$\xi(r)$ at $r  \lsim 100$ $\hkpc$ is quite flat  for the $\Vnow$ case
but continues to rise for the $\Vin$ case.  This reflects the enhanced
``destruction'' included in the $\Vnow$ model (galaxies in the centers
of halos tend  to lose mass quickly and their  $\Vmax$ values drop out
of the sample.)  If we characterize the clustering vs.  number density
trend by fitting  the amplitude of the $z=0$  correlation functions at
$100  \hkpc$ with  $\xi_{100} \propto  n_g^{-\alpha}$ we  find $\alpha
\simeq 0.15$ for $\Vnow$ and $\alpha \simeq 0.5$ for $\Vin$.  From the
above  arguments  with $\xi_{100}  \approx  \bar{\xi}$  we would  then
expect  $\Nc  \propto  n_g^{0.85}$  and $\Nc  \propto  n_g^{0.5}$  for
$\Vnow$ and $\Vin$ respectively.   Indeed, these are approximately the
slopes measured at $z=0$ in Figure~\ref{fig:Nc_n}.


\bibliography{paper7.bbl}

\end{document}